\documentclass[
aps,prd,amsmath,amssymb,superscriptaddress,nofootinbib,10pt,twocolumn]{revtex4-2}

\usepackage{xspace}
\usepackage{aas_macros}
\usepackage{graphicx}
\usepackage{graphics}
\usepackage[table,svgnames,dvipsnames]{xcolor}
\usepackage{hyperref}
\hypersetup{colorlinks=true,citecolor=blue,filecolor=blue,urlcolor=blue,linkcolor=blue}
\usepackage{orcidlink} 
\usepackage{hanging} 
\usepackage{verbatim}
\usepackage{amsmath}
\usepackage{multirow}

\usepackage{xurl}

\begin{document}

\title{Construction of the Damped Ly$\alpha$ Absorber Catalog for DESI DR2 Ly$\alpha$ BAO}


\author{A.~Brodzeller}
\affiliation{Lawrence Berkeley National Laboratory, 1 Cyclotron Road, Berkeley, CA 94720, USA}
\author{M.~Wolfson}
\affiliation{Center for Cosmology and AstroParticle Physics, The Ohio State University, 191 West Woodruff Avenue, Columbus, OH 43210, USA}
\affiliation{Department of Physics, The Ohio State University, 191 West Woodruff Avenue, Columbus, OH 43210, USA}
\affiliation{Department of Astronomy, The Ohio State University, 4055 McPherson Laboratory, 140 W 18th Avenue, Columbus, OH 43210, USA}
\author{D.~M.~Santos}
\affiliation{Aix Marseille Univ, CNRS, CNES, LAM, Marseille, France}
\author{M.~Ho}
\affiliation{University of Michigan, 500 S. State Street, Ann Arbor, MI 48109, USA}
\author{T.~Tan}
\affiliation{IRFU, CEA, Universit\'{e} Paris-Saclay, F-91191 Gif-sur-Yvette, France}
\author{M.~M.~Pieri}
\affiliation{Aix Marseille Univ, CNRS, CNES, LAM, Marseille, France}
\author{A.~Cuceu}
\affiliation{NASA Einstein Fellow}
\affiliation{Lawrence Berkeley National Laboratory, 1 Cyclotron Road, Berkeley, CA 94720, USA}

\author{M.~Abdul-Karim}
\affiliation{IRFU, CEA, Universit\'{e} Paris-Saclay, F-91191 Gif-sur-Yvette, France}
\author{J.~Aguilar}
\affiliation{Lawrence Berkeley National Laboratory, 1 Cyclotron Road, Berkeley, CA 94720, USA}
\author{S.~Ahlen}
\affiliation{Physics Dept., Boston University, 590 Commonwealth Avenue, Boston, MA 02215, USA}
\author{A.~Anand}
\affiliation{Lawrence Berkeley National Laboratory, 1 Cyclotron Road, Berkeley, CA 94720, USA}
\author{U.~Andrade}
\affiliation{Leinweber Center for Theoretical Physics, University of Michigan, 450 Church Street, Ann Arbor, Michigan 48109-1040, USA}
\affiliation{University of Michigan, 500 S. State Street, Ann Arbor, MI 48109, USA}
\author{E.~Armengaud}
\affiliation{IRFU, CEA, Universit\'{e} Paris-Saclay, F-91191 Gif-sur-Yvette, France}
\author{A.~Aviles}
\affiliation{Instituto Avanzado de Cosmolog\'{\i}a A.~C., San Marcos 11 - Atenas 202. Magdalena Contreras. Ciudad de M\'{e}xico C.~P.~10720, M\'{e}xico}
\affiliation{Instituto de Ciencias F\'{\i}sicas, Universidad Nacional Aut\'onoma de M\'exico, Av. Universidad s/n, Cuernavaca, Morelos, C.~P.~62210, M\'exico}
\author{S.~Bailey}
\affiliation{Lawrence Berkeley National Laboratory, 1 Cyclotron Road, Berkeley, CA 94720, USA}
\author{A.~Bault}
\affiliation{Lawrence Berkeley National Laboratory, 1 Cyclotron Road, Berkeley, CA 94720, USA}
\author{D.~Bianchi}
\affiliation{Dipartimento di Fisica ``Aldo Pontremoli'', Universit\`a degli Studi di Milano, Via Celoria 16, I-20133 Milano, Italy}
\affiliation{INAF-Osservatorio Astronomico di Brera, Via Brera 28, 20122 Milano, Italy}
\author{D.~Brooks}
\affiliation{Department of Physics \& Astronomy, University College London, Gower Street, London, WC1E 6BT, UK}
\author{R.~Canning}
\affiliation{Institute of Cosmology and Gravitation, University of Portsmouth, Dennis Sciama Building, Portsmouth, PO1 3FX, UK}
\author{L.~Casas}
\affiliation{Institut de F\'{i}sica d’Altes Energies (IFAE), The Barcelona Institute of Science and Technology, Edifici Cn, Campus UAB, 08193, Bellaterra (Barcelona), Spain}
\author{M.~Charles}
\affiliation{The Ohio State University, Columbus, 43210 OH, USA}
\author{E.~Chaussidon}
\affiliation{Lawrence Berkeley National Laboratory, 1 Cyclotron Road, Berkeley, CA 94720, USA}
\author{J.~Chaves-Montero}
\affiliation{Institut de F\'{i}sica d’Altes Energies (IFAE), The Barcelona Institute of Science and Technology, Edifici Cn, Campus UAB, 08193, Bellaterra (Barcelona), Spain}
\author{D.~Chebat}
\affiliation{IRFU, CEA, Universit\'{e} Paris-Saclay, F-91191 Gif-sur-Yvette, France}
\author{T.~Claybaugh}
\affiliation{Lawrence Berkeley National Laboratory, 1 Cyclotron Road, Berkeley, CA 94720, USA}
\author{K.~S.~Dawson}
\affiliation{Department of Physics and Astronomy, The University of Utah, 115 South 1400 East, Salt Lake City, UT 84112, USA}
\author{R.~de Belsunce}
\affiliation{Lawrence Berkeley National Laboratory, 1 Cyclotron Road, Berkeley, CA 94720, USA}
\author{A.~de la Macorra}
\affiliation{Instituto de F\'{\i}sica, Universidad Nacional Aut\'{o}noma de M\'{e}xico,  Circuito de la Investigaci\'{o}n Cient\'{\i}fica, Ciudad Universitaria, Cd. de M\'{e}xico  C.~P.~04510,  M\'{e}xico}
\author{A.~de~Mattia}
\affiliation{IRFU, CEA, Universit\'{e} Paris-Saclay, F-91191 Gif-sur-Yvette, France}
\author{Arjun~Dey}
\affiliation{NSF NOIRLab, 950 N. Cherry Ave., Tucson, AZ 85719, USA}
\author{Biprateep~Dey}
\affiliation{Department of Astronomy \& Astrophysics, University of Toronto, Toronto, ON M5S 3H4, Canada}
\affiliation{Department of Physics \& Astronomy and Pittsburgh Particle Physics, Astrophysics, and Cosmology Center (PITT PACC), University of Pittsburgh, 3941 O'Hara Street, Pittsburgh, PA 15260, USA}
\author{P.~Doel}
\affiliation{Department of Physics \& Astronomy, University College London, Gower Street, London, WC1E 6BT, UK}
\author{M.~Doshi}
\affiliation{University of California, Berkeley, 110 Sproul Hall \#5800 Berkeley, CA 94720, USA}
\affiliation{Lawrence Berkeley National Laboratory, 1 Cyclotron Road, Berkeley, CA 94720, USA}
\author{W.~Elbers}
\affiliation{Institute for Computational Cosmology, Department of Physics, Durham University, South Road, Durham DH1 3LE, UK}
\author{S.~Ferraro}
\affiliation{Lawrence Berkeley National Laboratory, 1 Cyclotron Road, Berkeley, CA 94720, USA}
\affiliation{University of California, Berkeley, 110 Sproul Hall \#5800 Berkeley, CA 94720, USA}
\author{A.~Font-Ribera}
\affiliation{Institut de F\'{i}sica d’Altes Energies (IFAE), The Barcelona Institute of Science and Technology, Edifici Cn, Campus UAB, 08193, Bellaterra (Barcelona), Spain}
\author{J.~E.~Forero-Romero}
\affiliation{Departamento de F\'isica, Universidad de los Andes, Cra. 1 No. 18A-10, Edificio Ip, CP 111711, Bogot\'a, Colombia}
\affiliation{Observatorio Astron\'omico, Universidad de los Andes, Cra. 1 No. 18A-10, Edificio H, CP 111711 Bogot\'a, Colombia}
\author{C.~Garcia-Quintero}
\affiliation{NASA Einstein Fellow}
\affiliation{Center for Astrophysics $|$ Harvard \& Smithsonian, 60 Garden Street, Cambridge, MA 02138, USA}
\author{L.~H.~Garrison}
\affiliation{Center for Computational Astrophysics, Flatiron Institute, 162 5\textsuperscript{th} Avenue, New York, NY 10010, USA}
\affiliation{Scientific Computing Core, Flatiron Institute, 162 5\textsuperscript{th} Avenue, New York, NY 10010, USA}
\author{E.~Gaztañaga}
\affiliation{Institut d'Estudis Espacials de Catalunya (IEEC), c/ Esteve Terradas 1, Edifici RDIT, Campus PMT-UPC, 08860 Castelldefels, Spain}
\affiliation{Institute of Cosmology and Gravitation, University of Portsmouth, Dennis Sciama Building, Portsmouth, PO1 3FX, UK}
\affiliation{Institute of Space Sciences, ICE-CSIC, Campus UAB, Carrer de Can Magrans s/n, 08913 Bellaterra, Barcelona, Spain}
\author{S.~Gontcho~A~Gontcho}
\affiliation{Lawrence Berkeley National Laboratory, 1 Cyclotron Road, Berkeley, CA 94720, USA}
\author{A.~X.~Gonzalez-Morales}
\affiliation{Departamento de F\'{\i}sica, DCI-Campus Le\'{o}n, Universidad de Guanajuato, Loma del Bosque 103, Le\'{o}n, Guanajuato C.~P.~37150, M\'{e}xico.}
\author{D.~Green}
\affiliation{Department of Physics and Astronomy, University of California, Irvine, 92697, USA}
\author{G.~Gutierrez}
\affiliation{Fermi National Accelerator Laboratory, PO Box 500, Batavia, IL 60510, USA}
\author{J.~Guy}
\affiliation{Lawrence Berkeley National Laboratory, 1 Cyclotron Road, Berkeley, CA 94720, USA}
\author{C.~Hahn}
\affiliation{Steward Observatory, University of Arizona, 933 N, Cherry Ave, Tucson, AZ 85721, USA}
\author{M.~Herbold}
\affiliation{The Ohio State University, Columbus, 43210 OH, USA}
\author{H.~K.~Herrera-Alcantar}
\affiliation{Institut d'Astrophysique de Paris. 98 bis boulevard Arago. 75014 Paris, France}
\affiliation{IRFU, CEA, Universit\'{e} Paris-Saclay, F-91191 Gif-sur-Yvette, France}
\author{K.~Honscheid}
\affiliation{Center for Cosmology and AstroParticle Physics, The Ohio State University, 191 West Woodruff Avenue, Columbus, OH 43210, USA}
\affiliation{Department of Physics, The Ohio State University, 191 West Woodruff Avenue, Columbus, OH 43210, USA}
\affiliation{The Ohio State University, Columbus, 43210 OH, USA}
\author{C.~Howlett}
\affiliation{School of Mathematics and Physics, University of Queensland, Brisbane, QLD 4072, Australia}
\author{D.~Huterer}
\affiliation{Department of Physics, University of Michigan, 450 Church Street, Ann Arbor, MI 48109, USA}
\affiliation{University of Michigan, 500 S. State Street, Ann Arbor, MI 48109, USA}
\author{M.~Ishak}
\affiliation{Department of Physics, The University of Texas at Dallas, 800 W. Campbell Rd., Richardson, TX 75080, USA}
\author{S.~Juneau}
\affiliation{NSF NOIRLab, 950 N. Cherry Ave., Tucson, AZ 85719, USA}
\author{R.~Kehoe}
\affiliation{Department of Physics, Southern Methodist University, 3215 Daniel Avenue, Dallas, TX 75275, USA}
\author{T.~Kisner}
\affiliation{Lawrence Berkeley National Laboratory, 1 Cyclotron Road, Berkeley, CA 94720, USA}
\author{A.~Kremin}
\affiliation{Lawrence Berkeley National Laboratory, 1 Cyclotron Road, Berkeley, CA 94720, USA}
\author{O.~Lahav}
\affiliation{Department of Physics \& Astronomy, University College London, Gower Street, London, WC1E 6BT, UK}
\author{C.~Lamman}
\affiliation{Center for Astrophysics $|$ Harvard \& Smithsonian, 60 Garden Street, Cambridge, MA 02138, USA}
\author{M.~Landriau}
\affiliation{Lawrence Berkeley National Laboratory, 1 Cyclotron Road, Berkeley, CA 94720, USA}
\author{J.M.~Le~Goff}
\affiliation{IRFU, CEA, Universit\'{e} Paris-Saclay, F-91191 Gif-sur-Yvette, France}
\author{L.~Le~Guillou}
\affiliation{Sorbonne Universit\'{e}, CNRS/IN2P3, Laboratoire de Physique Nucl\'{e}aire et de Hautes Energies (LPNHE), FR-75005 Paris, France}
\author{A.~Leauthaud}
\affiliation{Department of Astronomy and Astrophysics, UCO/Lick Observatory, University of California, 1156 High Street, Santa Cruz, CA 95064, USA}
\affiliation{Department of Astronomy and Astrophysics, University of California, Santa Cruz, 1156 High Street, Santa Cruz, CA 95065, USA}
\author{M.~E.~Levi}
\affiliation{Lawrence Berkeley National Laboratory, 1 Cyclotron Road, Berkeley, CA 94720, USA}
\author{Q.~Li}
\affiliation{Department of Physics and Astronomy, The University of Utah, 115 South 1400 East, Salt Lake City, UT 84112, USA}
\author{M.~Manera}
\affiliation{Departament de F\'{i}sica, Serra H\'{u}nter, Universitat Aut\`{o}noma de Barcelona, 08193 Bellaterra (Barcelona), Spain}
\affiliation{Institut de F\'{i}sica d’Altes Energies (IFAE), The Barcelona Institute of Science and Technology, Edifici Cn, Campus UAB, 08193, Bellaterra (Barcelona), Spain}
\author{P.~Martini}
\affiliation{Center for Cosmology and AstroParticle Physics, The Ohio State University, 191 West Woodruff Avenue, Columbus, OH 43210, USA}
\affiliation{Department of Astronomy, The Ohio State University, 4055 McPherson Laboratory, 140 W 18th Avenue, Columbus, OH 43210, USA}
\affiliation{The Ohio State University, Columbus, 43210 OH, USA}
\author{A.~Meisner}
\affiliation{NSF NOIRLab, 950 N. Cherry Ave., Tucson, AZ 85719, USA}
\author{J.~Mena-Fern\'andez}
\affiliation{Laboratoire de Physique Subatomique et de Cosmologie, 53 Avenue des Martyrs, 38000 Grenoble, France}
\author{R.~Miquel}
\affiliation{Instituci\'{o} Catalana de Recerca i Estudis Avan\c{c}ats, Passeig de Llu\'{\i}s Companys, 23, 08010 Barcelona, Spain}
\affiliation{Institut de F\'{i}sica d’Altes Energies (IFAE), The Barcelona Institute of Science and Technology, Edifici Cn, Campus UAB, 08193, Bellaterra (Barcelona), Spain}
\author{J.~Moustakas}
\affiliation{Department of Physics and Astronomy, Siena College, 515 Loudon Road, Loudonville, NY 12211, USA}
\author{A.~Muñoz-Gutiérrez}
\affiliation{Instituto de F\'{\i}sica, Universidad Nacional Aut\'{o}noma de M\'{e}xico,  Circuito de la Investigaci\'{o}n Cient\'{\i}fica, Ciudad Universitaria, Cd. de M\'{e}xico  C.~P.~04510,  M\'{e}xico}
\author{A.~D.~Myers}
\affiliation{Department of Physics \& Astronomy, University  of Wyoming, 1000 E. University, Dept.~3905, Laramie, WY 82071, USA}
\author{S.~Nadathur}
\affiliation{Institute of Cosmology and Gravitation, University of Portsmouth, Dennis Sciama Building, Portsmouth, PO1 3FX, UK}
\author{L.~Napolitano}
\affiliation{Department of Physics \& Astronomy, University  of Wyoming, 1000 E. University, Dept.~3905, Laramie, WY 82071, USA}
\author{H.~E.~Noriega}
\affiliation{Instituto de Ciencias F\'{\i}sicas, Universidad Nacional Aut\'onoma de M\'exico, Av. Universidad s/n, Cuernavaca, Morelos, C.~P.~62210, M\'exico}
\affiliation{Instituto de F\'{\i}sica, Universidad Nacional Aut\'{o}noma de M\'{e}xico,  Circuito de la Investigaci\'{o}n Cient\'{\i}fica, Ciudad Universitaria, Cd. de M\'{e}xico  C.~P.~04510,  M\'{e}xico}
\author{E.~Paillas}
\affiliation{Department of Physics and Astronomy, University of Waterloo, 200 University Ave W, Waterloo, ON N2L 3G1, Canada}
\affiliation{Steward Observatory, University of Arizona, 933 N, Cherry Ave, Tucson, AZ 85721, USA}
\affiliation{Waterloo Centre for Astrophysics, University of Waterloo, 200 University Ave W, Waterloo, ON N2L 3G1, Canada}
\author{N.~Palanque-Delabrouille}
\affiliation{IRFU, CEA, Universit\'{e} Paris-Saclay, F-91191 Gif-sur-Yvette, France}
\affiliation{Lawrence Berkeley National Laboratory, 1 Cyclotron Road, Berkeley, CA 94720, USA}
\author{W.~J.~Percival}
\affiliation{Department of Physics and Astronomy, University of Waterloo, 200 University Ave W, Waterloo, ON N2L 3G1, Canada}
\affiliation{Perimeter Institute for Theoretical Physics, 31 Caroline St. North, Waterloo, ON N2L 2Y5, Canada}
\affiliation{Waterloo Centre for Astrophysics, University of Waterloo, 200 University Ave W, Waterloo, ON N2L 3G1, Canada}
\author{C.~Poppett}
\affiliation{Lawrence Berkeley National Laboratory, 1 Cyclotron Road, Berkeley, CA 94720, USA}
\affiliation{Space Sciences Laboratory, University of California, Berkeley, 7 Gauss Way, Berkeley, CA  94720, USA}
\affiliation{University of California, Berkeley, 110 Sproul Hall \#5800 Berkeley, CA 94720, USA}
\author{F.~Prada}
\affiliation{Instituto de Astrof\'{i}sica de Andaluc\'{i}a (CSIC), Glorieta de la Astronom\'{i}a, s/n, E-18008 Granada, Spain}
\author{I.~P\'erez-R\`afols}
\affiliation{Departament de F\'isica, EEBE, Universitat Polit\`ecnica de Catalunya, c/Eduard Maristany 10, 08930 Barcelona, Spain}
\author{C.~Ram\'irez-P\'erez}
\affiliation{Institut de F\'{i}sica d’Altes Energies (IFAE), The Barcelona Institute of Science and Technology, Edifici Cn, Campus UAB, 08193, Bellaterra (Barcelona), Spain}
\author{C.~Ravoux}
\affiliation{Universit\'{e} Clermont-Auvergne, CNRS, LPCA, 63000 Clermont-Ferrand, France}
\author{J.~Rohlf}
\affiliation{Physics Dept., Boston University, 590 Commonwealth Avenue, Boston, MA 02215, USA}
\author{G.~Rossi}
\affiliation{Department of Physics and Astronomy, Sejong University, 209 Neungdong-ro, Gwangjin-gu, Seoul 05006, Republic of Korea}
\author{E.~Sanchez}
\affiliation{CIEMAT, Avenida Complutense 40, E-28040 Madrid, Spain}
\author{D.~Schlegel}
\affiliation{Lawrence Berkeley National Laboratory, 1 Cyclotron Road, Berkeley, CA 94720, USA}
\author{M.~Schubnell}
\affiliation{Department of Physics, University of Michigan, 450 Church Street, Ann Arbor, MI 48109, USA}
\affiliation{University of Michigan, 500 S. State Street, Ann Arbor, MI 48109, USA}
\author{F.~Sinigaglia}
\affiliation{Departamento de Astrof\'{\i}sica, Universidad de La Laguna (ULL), E-38206, La Laguna, Tenerife, Spain}
\affiliation{Instituto de Astrof\'{\i}sica de Canarias, C/ V\'{\i}a L\'{a}ctea, s/n, E-38205 La Laguna, Tenerife, Spain}
\author{D.~Sprayberry}
\affiliation{NSF NOIRLab, 950 N. Cherry Ave., Tucson, AZ 85719, USA}
\author{G.~Tarl\'{e}}
\affiliation{University of Michigan, 500 S. State Street, Ann Arbor, MI 48109, USA}
\author{P.~Taylor}
\affiliation{The Ohio State University, Columbus, 43210 OH, USA}
\author{W.~Turner}
\affiliation{Center for Cosmology and AstroParticle Physics, The Ohio State University, 191 West Woodruff Avenue, Columbus, OH 43210, USA}
\affiliation{Department of Astronomy, The Ohio State University, 4055 McPherson Laboratory, 140 W 18th Avenue, Columbus, OH 43210, USA}
\affiliation{The Ohio State University, Columbus, 43210 OH, USA}
\author{M.~Walther}
\affiliation{Excellence Cluster ORIGINS, Boltzmannstrasse 2, D-85748 Garching, Germany}
\affiliation{University Observatory, Faculty of Physics, Ludwig-Maximilians-Universit\"{a}t, Scheinerstr. 1, 81677 M\"{u}nchen, Germany}
\author{B.~A.~Weaver}
\affiliation{NSF NOIRLab, 950 N. Cherry Ave., Tucson, AZ 85719, USA}
\author{C.~Yèche}
\affiliation{IRFU, CEA, Universit\'{e} Paris-Saclay, F-91191 Gif-sur-Yvette, France}
\author{R.~Zhou}
\affiliation{Lawrence Berkeley National Laboratory, 1 Cyclotron Road, Berkeley, CA 94720, USA}
\author{H.~Zou}
\affiliation{National Astronomical Observatories, Chinese Academy of Sciences, A20 Datun Rd., Chaoyang District, Beijing, 100012, P.R. China}
\author{S.~Zou}
\affiliation{Department of Astronomy, Tsinghua University, 30 Shuangqing Road, Haidian District, Beijing, China, 100190}

\collaboration{DESI Collaboration}

\begin{abstract}

We present the Damped Ly$\alpha$ Toolkit for automated detection and characterization of Damped Ly$\alpha$ absorbers (DLA) in quasar spectra. Our method uses quasar spectral templates with and without absorption from intervening DLAs to reconstruct observed quasar forest regions. The best-fitting model determines whether a DLA is present while estimating the redshift and \texttt{HI} column density. With an optimized quality cut on detection significance ($\Delta \chi_{r}^2>0.03$), the technique achieves an estimated 80\% purity and 79\% completeness when evaluated on simulated spectra with S/N~$>2$ that are free of broad absorption lines (BAL). We provide a catalog containing candidate DLAs from the DLA Toolkit detected in DESI DR1 quasar spectra, of which 21,719 were found in S/N~$>2$ spectra with predicted $\log_{10} (N_\texttt{HI}) > 20.3$ and detection significance $\Delta \chi_{r}^2 >0.03$. We compare the Damped Ly$\alpha$ Toolkit to two alternative DLA finders based on a convolutional neural network (CNN) and Gaussian process (GP) models. We present a strategy for combining these three techniques to produce a high-fidelity DLA catalog from DESI DR2 for the Ly$\alpha$ forest baryon acoustic oscillation measurement. The combined catalog contains 41,152 candidate DLAs with $\log_{10} (N_\texttt{HI}) > 20.3$ from quasar spectra with S/N~$>2$. We estimate this sample to be approximately 85\% pure and 79\% complete when BAL quasars are excluded.

\end{abstract}

\maketitle

\section{Introduction}
\label{sect:intro}
Damped Ly$\alpha$ absorbers (DLA) form a class of quasar absorption systems caused by foreground neutral hydrogen reservoirs with column densities $N_\texttt{HI}>2 \times 10^{20}$ cm$^{-2}$ \citep{wolfe86,wolfe05}. This column density is sufficient to be almost entirely neutral, creating a system that is self-shielded against the ionizing background \citep[][]{vladilo01}. As the dominant source of neutral hydrogen in the universe, DLAs provide a valuable insight into galaxy formation history and evolution \citep[e.g.][]{prochaska97,pontzen08,noterdaeme09, kulkarni22}. They probe the physical conditions of their associated galaxies, such as the star formation rate, metal content, and halo mass \citep[e.g.][]{cen12,fontribera12b,krogager13,fumagalli14,neeleman18,krogager20,oyazun24,dharmender24}. 

DLAs also play a role in cosmological measurements. They serve as tracers of the matter density field \citep{fontribera12b,perez18a}, from which the baryon acoustic oscillation scale can be measured \citep{perez23}. While valuable as tracers, they are contaminants for the more powerful density tracer: the Ly$\alpha$ forest. If not properly accounted for, the presence of a DLA can impact the quasar mean continuum estimate and bias the extracted neutral hydrogen density field. 
The broad damping wings of their absorption profile introduce additional correlations and noise to the 3D Ly$\alpha$ forest correlation function, which can substantially impact the inferred redshift space distortion parameter and linear bias parameter \citep[e.g.][]{slosar11,fontribera12a,rogers18a,rogers18b, wang22,tan25}. Further, unidentified DLAs add power on large scales to the 1D power spectrum, shifting the value of the measured scalar spectral index \citep[e.g.][]{mcdonald05,chabanier19,karacayli24}. For these reasons, in addition to their value as astrophysical probes, significant efforts have been made to efficiently detect and characterize DLAs in high-redshift quasar surveys over the past several decades.

\citet{prochaska04} and \citet{prochaska05} led semi-automated searches for DLAs in the early Sloan Digital Sky Survey \citep[SDSS;][]{york_SDSS}. They identified DLA candidates by sliding a window over the Ly$\alpha$ forest to find regions with signal-to-noise ratios (S/N) significantly lower than the characteristic S/N of the quasar. Such regions were visually inspected to confirm detection and $N_\texttt{HI}$ was estimated with a by-eye Voigt profile fit to the trough.

As the SDSS quasar sample rapidly grew, fully automated pipelines for surveying DLAs became necessary. The technique by \citet{noterdaeme09,noterdaeme12} correlated observed quasar spectra with synthetic Voigt profiles over the plausible ($N_\texttt{HI}$, z)-surface, using a metal absorption template to refine DLA redshift solutions. The authors demonstrated performance comparable to that in \citep{prochaska04} without requiring a human intervention step. 

\citet{garnett17} proposed a DLA pipeline based on Gaussian process (GP) models for the quasar emission spectrum with and without intervening DLAs. This method, later improved by \citet{ho20,ho21}, returns the probability that a given quasar sightline contains up to 3 DLAs using Bayesian model selection with a prior on the column density distribution function informed by previous DLA surveys. The choice to simultaneously model the quasar flux and DLA profile aimed to reduce the false detection rate from incomplete or insufficient quasar emission functions. The runtime of this method also scaled efficiently with sample size, an important asset in an era of large galaxy surveys. \citet{wang22} reported the GP method achieved sample completeness and purity of more than 88\% for absorbers with $\log N_\texttt{HI} >20.0$ when evaluated on mock spectra of the first year of the Dark Energy Spectroscopic Instrument (DESI) survey with a median S/N~$>3$ over $1420-1480$~\AA\ that were free of broad absorption lines (BALs).

Motivated by the human expert's ability to identify the DLA signatures in quasar forests, \citet{parks18} tasked a convolutional neural network (CNN) with characterizing an arbitrary number of DLAs on quasar sightlines. They trained the CNN on artificial DLA profiles injected into real DLA-free spectra and demonstrated it could recover $\sim 80$\% of the DLAs reported by the DLA survey of \citet{noterdaeme09}. \citet{chabanier22} independently validated the CNN, finding excellent efficiency and purity for bright sources but biased $N_\texttt{HI}$ estimates. The estimates were particularly skewed high in the low $N_\texttt{HI}$ regime, though this could be mitigated with cuts on the CNN's detection confidence output or removed with a post hoc Voigt profile fit at the CNN predicted redshift. \citet{wang22} retrained the CNN on \texttt{Ly$\alpha$CoLoRe} mock spectra representative of the DESI year one quasar sample \citep[][]{farr20_lyacolore}. They reported purity and completeness values mostly over 90\% on a BAL-free mock spectra sample with S/N~$>3$ for absorbers with $\log N_\texttt{HI} >20.0$. They do not report a bias on $N_\texttt{HI}$ from their work.

Both the GP and CNN DLA finders were used to construct a concordance catalog with the first data release (DR1) sample from the DESI survey \citep{DESI_DR1}. DLAs in the concordance catalog\footnote{\url{https://data.desi.lbl.gov/public/dr1/vac/dr1/dla-cnn-gp}} are detected by both algorithms with redshift solutions within $800$~km~s$^{-1}$. There is no requirement that the column density estimates agree to account for biases on $N_\texttt{HI}$ from the CNN, and the GP solution for $N_\texttt{HI}$ and redshift was assumed for all detections. This catalog informed contaminant masking in the DESI DR1 Ly$\alpha$ forest BAO measurement \cite[][]{DESI-Y1KP6}. The BAO analysis required a detection probability greater than 50\% from both finders and forest S/N~$>3$ to further boost the catalog's purity. 

In this paper, we present the Damped Ly$\alpha$ Toolkit for DLA detection and characterization. The DLA Toolkit,  is based on spectral template fitting designed to simultaneously model the quasar flux and up to 3 DLA absorption profiles per sightline. The best fitting model informs whether a DLA is present and, if so, its most likely redshift and $N_\texttt{HI}$. We demonstrate that the DLA Toolkit provides more than a $15$\% gain in completeness while maintaining an estimated purity level above 80\% when included in the combined catalog approach for Ly$\alpha$ BAO measurements with DESI DR2.  We also show that DLAs can be reliably classified in DESI spectra down to S/N~$=2$ for the background quasar.

This paper supports the DESI DR2 Ly$\alpha$ BAO measurement \citep{DESI_lyaBAO} and the subsequent cosmological interpretation when combined with BAO from galaxy samples \citep{DESI_galaxyBAO}. The DLA catalog presented in this work is a critical component of the Ly$\alpha$ BAO measurement, informing which spectral regions should be masked to ensure a robust measurement.
In Section~\ref{sect:data}, we review the simulated data used to validate the DLA Toolkit and the DESI quasar samples for which we produce DLA catalogs. 
We provide a complete description of the technique behind the DLA Toolkit in Section~\ref{sect:method}. Section~\ref{sect:performance} presents our validation study. We perform identical tests with the CNN and GP DLA finders for comparison. Section~\ref{sect:catalogs} describes the DLA catalog produced by the DLA Toolkit with DESI DR2 and the subset catalog corresponding to DR1, the latter of which will be made available with this paper.\footnote{\url{https://data.desi.lbl.gov/public/dr1/vac/dr1/dla-toolkit}} We then discuss optimal combinations of the GP, CNN, and DLA Toolkit catalogs from DESI DR2 to produce a high-fidelity catalog for the Ly$\alpha$ forest BAO measurement in Section~\ref{BAO_combined_catalog}. Given the combined catalog's importance, we validate its purity in real data using a stacking method here. We further compare the new strategy for the combined DLA catalog to that adopted for the DR1 BAO measurement. The impact of the DLA catalog and masking strategy on BAO is explored in a companion paper \citep{casas_mocks}. In Section~\ref{sect:future}, we discuss potential improvements to the DLA Toolkit and the combined catalog strategy with respect to other Ly$\alpha$ forest analyses. We conclude in Section~\ref{sect:discussion}.

\section{Data Samples}
\label{sect:data}
This section presents a brief overview of the DESI instrument and survey. We then describe the quasar samples from the DR1 and DR2 used in this work. Lastly, we introduce the synthetic quasar spectra used to validate the performance of the DLA Toolkit.

\subsection{The DESI Survey}

DESI is a multi-object, fiber-fed spectrograph installed on the Mayall 4-m telescope at Kitt Peak National Observatory \citep{DESI13,DESI16a,DESI16b}. DESI consists of a new 3.2-degree diameter prime focus corrector and a focal plane hosting 5000 robotic positioners with optical fibers that direct light from survey targets to 10 spectrographs \citep{silber23,miller24,poppett24}. The target selection is based on imaging from the DESI Legacy Imaging Survey \citep{dey19} and was extensively validated in the early survey \citep{DESI_SV}. In particular relevance to this paper is the quasar target selection discussed by \citet{chaussidon23}.  A complete overview of the DESI instrumentation is provided by \cite{desi22}, while the survey operations and strategy are reviewed by \cite{schlafly23}. 

We use DESI DR1 and DR2 in this work. DR1 consists of the spectroscopic data collected from approximately $14.5$ million extragalactic objects and $4$ million stars during the first year of main survey operations. This unprecedented data set, which includes reprocessing of the previous ``Early Data Release'' \citep{DESI_EDR}, enabled a range of key science papers presenting large-scale structure catalogs \citep{DESI-Y1KP3}, BAO measurements \citep{DESI-Y1KP6,DESI-Y1KP5}, full-shape clustering analyses \citep{DESI-Y1KP4}, and the cosmological implication of these measurements \citep{DESI-Y1KP7a,DESI-Y1KP7b}. DR2 is a superset of DR1, consisting of the spectroscopic data collected from approximately $33$ million extragalactic objects and $12$ million stars during the first three years of main survey operations. The data was reprocessed with the latest version of DESI's spectroscopic pipeline \citep{guy23} that features improved calibration procedures relative to DR1.

\subsection{DESI Quasar Samples}
\label{subsec:desi_data}

This work uses the same quasar redshift catalogs as DESI's Ly$\alpha$ forest BAO analyses \citep{DESI-Y1KP6,DESI_lyaBAO}. These catalogs are constructed following the logic presented by \cite{chaussidon23}, combining the results from several spectral classifiers to achieve a highly complete and pure quasar sample \citep{busca18,farr20_qn,brodzeller23,anand2024,green_qn}. The DR1 quasar redshifts are refined\footnote{The refined redshifts are available in the \texttt{zlya} value-added catalog: \url{https://data.desi.lbl.gov/public/dr1/vac/dr1/zlya}} from the standard quasar-classification pipeline redshifts to correct for a bias reported at $z>2$ \citep{wu23,bault24}. This bias has since been mitigated in the main pipeline and thus the correction is unnecessary in DR2. The quasar catalogs also include BAL attributes associated with C~\textsc{iv} and Si~\textsc{iv} features \citep[see][for information on BAL detection in DESI]{filbert24}. 

\subsection{Simulated Quasar Spectra}
\label{subsec:mock_data}

We validate the DLA Toolkit on synthetic realizations of the DESI DR2 Ly$\alpha$ quasar sample. The process for generating the synthetic data set closely follows that outlined by \cite{herrera23} with specifics regarding the DR2 realizations discussed by \cite{casas_mocks}. In particular, our validation study uses one realization of \texttt{Ly$\alpha$CoLoRe} mocks \citep{farr20_lyacolore,perez22} and one realization of the Saclay mocks\footnote{The Saclay mock spectra are generated using \texttt{SaclayMocks} package available at \url{https://github.com/igmhub/SaclayMocks}} \citep{etourneau24}. The Saclay and \texttt{Ly$\alpha$CoLoRe} mocks are generated following similar processes with key differences in how quasars populate the simulated density field and the prescription for adding redshift-space distortions to the forest. We observe comparable performance on both mock data samples, so we only present the \texttt{Ly$\alpha$CoLoRe} results for brevity.

Briefly, the \texttt{Ly$\alpha$CoLoRe} mocks use a matter density distribution simulated from Gaussian random fields. Quasar positions are drawn from Poisson sampling of this density distribution with an overdensity threshold criterion imposed. Transmitted flux skewers for each quasar sightline are then created from a fluctuating Gunn-Peterson approximation of a log-normal transformation of the density field and a velocity field determined by its Newtonian potential. These skewers are processed into realistic Ly$\alpha$ transmitted flux skewers by adding to each sightline redshift space distortions based on the velocity field and a one-dimensional Gaussian random field to account for small-scale fluctuations. Then, we use the \texttt{quickquasars} script from the \texttt{desisim} repository\footnote{\url{https://github.com/desihub/desisim}} to generate a sample of realistic synthetic quasar spectra that mimics the DESI DR2 footprint, redshift distribution, and magnitude distribution. At this stage, \texttt{quickquasars} post-processes the skewers by adding a quasar continuum template~\citep{simqso:2021}, instrumental noise~\citep{Kirkby:2016}, and absorption features due to IGM metal lines, BALs, and DLAs. BALs are randomly added to 16\% of the population following precomputed templates~\citep{martini24}. DLA positions and column densities are drawn from the same initial density field as the quasar positions and introduced into the spectra following a Voigt profile model.\footnote{More specifically, we add an absorption feature at $\lambda_{\rm obs} = \lambda_{\rm Ly\alpha}(z_{\rm HCD}+1)$ with $F_{\rm HCD} = \exp(-\tau_{\rm HCD})$ transmission. Here $\tau_{\rm HCD} = N_\texttt{HI}\sigma(\lambda_{\rm obs};\lambda,f,b,\Gamma)$ follows a Voigt-profile cross-section model parameterized by $\lambda_{\rm Ly\alpha} = 1215.67$ \AA\ wavelength, $f_{\rm Ly\alpha}=0.4164$ oscillator strength, $b=30\ {\rm km/s}$ Doppler width, and $\Gamma = 6.265\times10^8\ {\rm s^{-1}}$ spontaneous emission coefficient.} We use the $z$ and $N_\texttt{HI}$ truth values of the input DLAs to evaluate the ability of the DLA Toolkit to recover DLAs accurately.

\section{The DLA Toolkit}
\label{sect:method}
\subsection{Detection Method}
\label{subsec:dlatoolkit}

Our technique fundamentally relies on the fact that a quasar's spectrum and the absorption profile from an intervening DLA are unrelated and therefore separable. Assuming the quasar redshift is known, we can model an observed spectrum $f_{obs}$ as a product of the quasar flux $F_q$ and the Lyman series transmission vectors $T_{DLA,j\in(0,n]}$ for $n$ intervening DLAs with redshifts $z_j$ and column densities $N_{\texttt{HI},j}$ following Equation~\eqref{eq:f_obs}. 

\begin{equation}
\label{eq:f_obs}
    M(\lambda_{obs}) = \begin{cases} 
    n=0, & F_{q}(\lambda_{obs}) \\
    n>0, & \prod_{j=1}^{n} T_{DLA,j}(\lambda_{obs},z_{j},N_{\texttt{HI},j}) \\
     & \hspace{78pt} \times F_{q}(\lambda_{obs}) \\
    
    \end{cases}
\end{equation}

We assume a Voigt profile for $T_{DLA}$ that includes absorption from the Ly$\alpha$ and Ly$\beta$ transitions. For $F_q$, we use version 1.1 of the \texttt{HIZ} quasar templates \citep{brodzeller23} from the redshift fitting software \texttt{Redrock}.\footnote{\url{https://github.com/desihub/redrock}}\textsuperscript{,}\footnote{\url{https://github.com/desihub/redrock-templates}} These templates consist of 4 eigenspectra derived using approximately 140,000 SDSS  $z_{QSO}>1.3$ quasar spectra. The eigenspectra incorporate the Ly$\alpha$ effective optical depth model $\tau_{\text{eff}}$ from \cite{kamble20}. $F_q$ is thus defined in Equation~\eqref{eq:fq_pca}, where $v_i$ is the ith eigenspectrum and $a_i$ is the coefficient on that eigenspectrum that provides the optimal reconstruction. 

\begin{equation}
\label{eq:fq_pca}
    F_q (\lambda_{obs}) = e^{-\tau_{\text{eff}}(\lambda_{obs})} \sum_i^{N=4} a_i v_i (\lambda_{obs})
\end{equation}

The procedure for detecting DLAs on a given quasar sightline is as follows. First, we define the redshift boundaries of the DLA search window using Equation~\eqref{eq:zmin} and Equation~\eqref{eq:zmax}, following \cite{ho20}. The minimum DLA redshift avoids wavelengths blueward of the Lyman limit with a buffer for potential error on quasar redshift $z_{QSO}$. The maximum DLA redshift mitigates the risk that the absorption profile will be used to compensate the quasar flux model for peculiarities in the Ly$\alpha$ emission line, such as asymmetry or strong intrinsic absorption. The allowed \texttt{HI} column density range is  $20.1 < \log_{10}(N_\texttt{HI}) <22.6$. The minimum column density is lower than the canonical $\log_{10} (N_{\texttt{HI}}) = 20.3$ definition for DLAs to avoid false positives from sub-DLAs. Detections with a column density below the DLA threshold can be removed in post-processing if desired.

\begin{table*}[htb]

\centering
\caption{DLA Catalog Columns}

\begin{tabular}{|c | c | c |} 
\hline
Column & Type & Description \\
 \hline
TARGETID & int64 & unique DESI target identifier for quasar \\
RA & double & right ascension in decimal degrees (J2000) \\ 
DEC & double & declination in decimal degrees (J2000) \\
Z\_QSO & double & quasar redshift \\ 
SNR\_FOREST & double & mean pixel S/N over $1040-1205$~\AA\ in Z\_QSO rest frame \\
SNR\_REDSIDE & double & mean pixel S/N over $1420-1480$~\AA\ in Z\_QSO rest frame \\
DLAID & char[20] & unique identifier for DLA \\
Z\_DLA & double & DLA redshift \\
Z\_DLA\_ERR & double & error on Z\_DLA estimated with parabola fit to $\chi^2$ minimum \\
NHI & double & log10 HI column density of DLA \\
NHI\_ERR & double & error on NHI estimated with parabola fit to $\chi^2$ minimum \\
COEFF & double[4] & coefficients on quasar eigenspectra for DLA solution\\
DELTACHI2 & double & improvement in reduced $\chi^2$ from including DLA \\
DLAFLAG\footnote{This column is absent from the DR1 DLA catalog since non-zero entries are discarded} & int64 & mask bit indicating potentially problematic fit \\
\hline
\end{tabular}

\label{tab:cat_columns}
\end{table*}

\begin{equation}
\label{eq:zmin}
z_{min} = \max 
    \begin{cases}
     \frac{912}{\lambda_{\text{Ly}\alpha}} (1 + z_{QSO}) - 1 + \frac{3000~\text{km}~\text{s}^{-1}}{c} & \\ 
    \frac{\min\lambda_{obs}}{\lambda_{\text{Ly}\alpha}} - 1 &
    \end{cases}
\end{equation}

\begin{equation}
\label{eq:zmax}
z_{max} = z_{QSO} - \frac{3000~\text{km}~\text{s}^{-1}}{c}
\end{equation}

If there are known C~\textsc{iv} BALs in the observed spectrum, we optionally mask the impacted regions. The velocity profile of C~\textsc{iv} BALs is extrapolated to mask for potential BALs from Si~\textsc{iv}, N~\textsc{iv}, and Ly$\alpha$ which regularly are co-occurring. The BAL masking strategy from \cite{ennesser22} is then applied to the forest region to help mitigate BAL/DLA confusion. We do not proceed with DLA detection if BAL masking results in a loss of more than $80$\% of the DLA search window. For the present work, BAL masking is always applied when BAL information is available.

Next, we shift the quasar templates to the observer frame using the provided $z_{QSO}$ and resample it to match the $\Delta \lambda$ binning of the observed spectrum. A null model, $n=0$ in Equation~\eqref{eq:f_obs}, is fit to the spectrum at $\lambda_{obs} > \max(\min(\lambda_{obs}), (1+z_{QSO}) \times 912$~\AA) by minimizing the reduced $\chi^2$ statistic defined in Equation~\eqref{eq:chi2}. There is no upper wavelength bound, as we wish to exploit the correlations between spectral features at $\lambda_{RF,QSO} < \lambda_{\text{Ly}\alpha}$ and $\lambda_{RF,QSO} > \lambda_{\text{Ly}\alpha}$ \citep[e.g.][]{paris11}.

\begin{equation}
\label{eq:chi2}
\begin{split}
    \chi_{r,n}^2 = \frac{1}{N-2n} \sum_{i}^N \frac{(f_{obs} (\lambda_{i}) - M(\lambda_i))^2}{\sigma_{pipe}^2(\lambda_i) + (M(\lambda_i)\sigma_{LSS}(\lambda_i))^2} 
\end{split}
\end{equation}

$M(\lambda_i)$ is defined by Equation~\eqref{eq:f_obs}, with the quasar eigenspectra coefficients being free parameters. The $\chi_r^2$ denominator consists of the flux variance estimated by the spectral reduction pipeline $\sigma_{pipe}^2$ and the intrinsic variance of the Lyman series flux transmission field $\sigma_{LSS}^2$ scaled by the squared estimate for observed flux. The $\sigma_{LSS}^2$ function in this work is set by the continuum fitting analysis described in \cite{DESI_lyaBAO} applied to an early version of the DESI DR2 sample (internally referred to as \texttt{jura}).\footnote{The DR1 equivalent, internally referred to as \texttt{iron}, for the $\sigma_{LSS}^2$ function will be available with the code. $\sigma_{LSS}^2$ from \texttt{jura} will be available no earlier than the publication of DESI DR2. The difference between $\sigma_{LSS}^2$ versions is minimal for the wavelength region concerned in this work.} This term only impacts $\lambda_{RF,QSO} < \lambda_{\text{Ly}\alpha}$ and is set to zero elsewhere.  

We save the reduced $\chi^2$ over the DLA search window of the best null fit. Next, we fit the spectrum with a 1-DLA model ($n=1$) over a coarse grid of the allowed $N_{\texttt{HI}}$ and $z$ ranges defined by steps of $\Delta \log_{10} N_\texttt{HI} = 0.05$ and $\Delta z = 0.01$. The eigenspectra coefficients are re-optimized at each grid point via Equation~\eqref{eq:chi2}, and the $\chi_r^2$ over the DLA search window is recorded. Resolving for the quasar eigenspectra coefficients at each step is crucial because if a DLA is indeed present it will have biased the null model towards underestimated flux. We then identify the best-fitting ($N_\texttt{HI}$, $z$)-pair and refit in finer steps spanning $\pm 0.02$ in $z$ and $\pm 0.15$ in $\log_{10}(N_\texttt{HI})$ about the minimum. We solve for the final column density and redshift estimates via a parabola fit to the refined $\chi_r^2$-surface. A parabola is iterativley fit in each dimension until convergence. Poor parabola fits, such as boundary relaxations, are flagged. A sparse visual inspection reveals most failed parabola fits originate from apparently DLA-free spectra for which we do not expect the $\chi_r^2$ surface to be well-defined by parabola. Flagged ``detections'' are maintained in the code's raw output; however, we choose to discard them from our final catalogs.

\begin{figure*}
\centering
	\includegraphics[width=\textwidth]{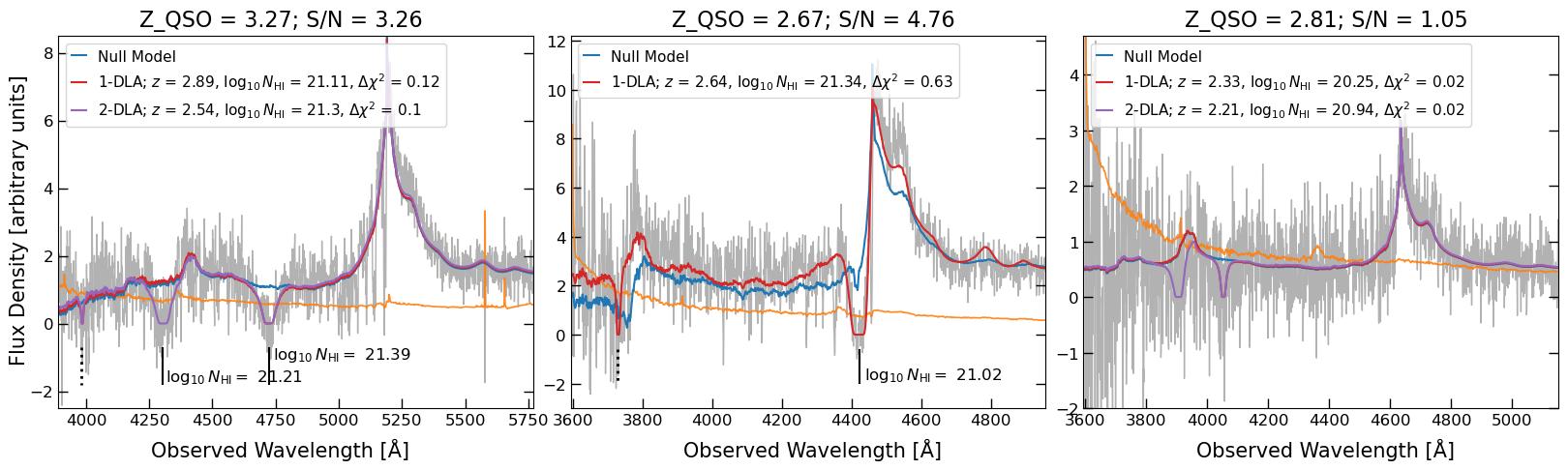}
    \caption{
        Three \texttt{Ly$\alpha$CoLoRe} quasar spectra (gray lines) and their simulated error spectra (orange lines) with the best fitting null model and n-DLA models that triggered a detection. The estimated $z$ and $N_\texttt{HI}$ of each detection is provided in the legends along with the $\Delta \chi_r^2$ of the fit. The solid vertical lines mark the location of true DLA troughs with the input $N_\texttt{HI}$ annotated. The dashed vertical lines indicate the corresponding true Ly$\beta$ troughs, if visible. 
        }
    \label{fig:example_detections}
\end{figure*}

A DLA detection is defined using the threshold parameter $\Delta \chi_r^2$ in Equation~\eqref{eq:detection_def}. This term quantifies how much the fit improves (or degrades) from including a DLA absorption profile. It is always computed using the reduced $\chi^2$ to account for the extra parameters in an $n$-DLA model relative to a $(n-1)$-DLA model.
To be maximally inclusive, the DLA Toolkit sets a weak threshold of $\Delta \chi_{r}^2 > 0.01$ to constitute a detection. In Section~\ref{subsect:mock_validation}, we investigate alternative choices for the $\Delta \chi_r^2$ detection threshold.

\begin{equation}
\label{eq:detection_def}
    \Delta \chi_r^2 = \chi^2_{r,n-1} - \chi^2_{r,n}
\end{equation}

If the $\Delta \chi_r^2$ for the 1-DLA model meets the detection threshold, we repeat the above procedure for a 2-DLA model and similarly for a 3-DLA model if merited. The solutions for any previously identified DLAs are fixed when solving subsequent models. The $\Delta \chi_r^2$ threshold for detection is uniform across all $n$. The DLA Toolkit stores the relevant information for each detection in an output catalog summarized in Table~\ref{tab:cat_columns}. 

Figure~\ref{fig:example_detections} shows three simulated spectra in which the DLA Toolkit detected a candidate DLA. The first two panels show true positive detections while the last panel is an example false positive. Many false positives originate in low S/N spectra and from true high column density absorbers but with $N_\texttt{HI}$ below the search limit of the DLA Toolkit. The first false detection in the last panel of Figure~\ref{fig:example_detections} aligns with the position of a true absorber with input $z=2.33$ and $\log_{10} (N_\texttt{HI}) = 19.75$. As discussed in Section~\ref{subsect:mock_validation}, the false positive rate can be mitigated with cuts on either S/N or $N_\texttt{HI}$.

\begin{figure}[ht]
\centering
    \includegraphics[width=\columnwidth]{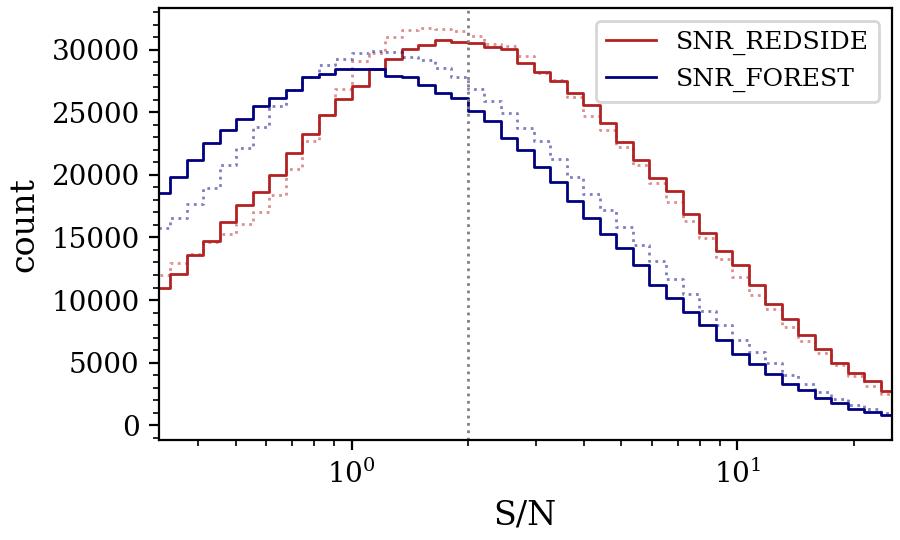}
    \caption{ Histogram illustrating the \texttt{SNR\_REDSIDE} and \texttt{SNR\_FOREST} distributions for the DESI DR2 quasar (solid lines) sample and the \texttt{Ly$\alpha$CoLoRe} mock sample (dotted lines). The vertical dashed line indicates S/N~$>2$.}
    \label{fig:snr_hist}
\end{figure}

\subsubsection{S/N Metrics}

The DLA Toolkit computes two S/N metrics that are saved in the code's output.
The first is \texttt{SNR\_REDSIDE} which is defined by the average S/N per pixel over $1420 \text{\AA}<\lambda_{RF,QSO}<1480 \text{\AA}$. 
The second is \texttt{SNR\_FOREST} which is defined by the average S/N per pixel over $1040 \text{\AA}<\lambda_{RF,QSO}<1205 \text{\AA}$. 
A value of $-1$ reflects insufficient wavelength coverage and is typically limited to \texttt{SNR\_FOREST} for lower redshift quasars. The S/N distributions for the DESI DR2 quasar sample and the \texttt{Ly$\alpha$CoLoRe} mock data sample are shown in Figure~\ref{fig:snr_hist}. As evidenced in the figure, \texttt{SNR\_REDSIDE} nearly always exceeds that computed in the forest by a factor of $1.5-2$, on average.
This trend is exaggerated when a DLA is present \citep{prochaska04,prochaska05}. 
We exclusively use \texttt{SNR\_REDSIDE} for any S/N value reported throughout this paper to avoid biasing against sightlines with DLAs.

\section{Performance}
\label{sect:performance}
We evaluate the DLA Toolkit on the sample of simulated spectra described in Section~\ref{subsec:mock_data} restricted to $2.0<z_{QSO}<4.25$. The lower limit ensures the Ly$\alpha$ forest is redshifted to the wavelength coverage of the DESI spectrographs while the upper limit comes from the the maximum quasar redshift used in the Ly$\alpha$ forest BAO measurement \cite{DESI-Y1KP6,DESI_lyaBAO}. Using the truth values for DLAs input into the mock spectra, we measure the column density and redshift accuracy of true positives. We then report on the purity and completeness of the resulting DLA sample and the dependence of these metrics on various parameters such as S/N. We perform identical tests with the GP and CNN DLA detection methods for comparison.

A persistent issue with DLA detection is BAL/DLA confusion. Since BAL contamination is removed before extracting Ly$\alpha$ flux transmission fields for BAO measurement \citep[see the DESI BAL masking strategy presented by][]{martini24}, we focus our validation on the BAL-free subset of the mock spectra sample. We include some discussion of performance on the full mock data sample, but all reported metrics and figures correspond to the BAL-free subsample unless explicitly stated otherwise. 

\subsection{Validation on Simulated Spectra}
\label{subsect:mock_validation}

We run the DLA Toolkit on the sample of approximately 930,000 simulated spectra that satisfy the quasar redshift restriction, of which $778,043$ are free of BALs. We trim the DLA Toolkit output catalog on predicted $\log_{10} (N_\texttt{HI})>20.3$ and remove flagged detections (see Section~\ref{subsec:dlatoolkit}). Approximately 13\% of quasar sightlines have a DLA detection after applying these cuts, with $\sim$1\% having multiple DLA detections.

The DLA truth catalog is defined as input DLAs with $\log_{10} (N_\texttt{HI}) > 20.3$ and redshifts within the wavelength search window defined by Equation~\eqref{eq:zmin} and Equation~\eqref{eq:zmax}. A detection by the DLA Toolkit is considered a true positive if Equation~\eqref{eq:z_req} is uniquely satisfied for any DLA in the truth catalog. This is equivalent to the predicted DLA being within 3,000 km~s$^{-1}$ of a true DLA. 

\begin{equation}
\left| \frac{z_{\text{predicted}} - z_{\text{true}}}{1 + z_{\text{true}}} \right| < 0.01 
\label{eq:z_req}
\end{equation}

\begin{figure}[t]
\centering
    \includegraphics[width=\columnwidth]{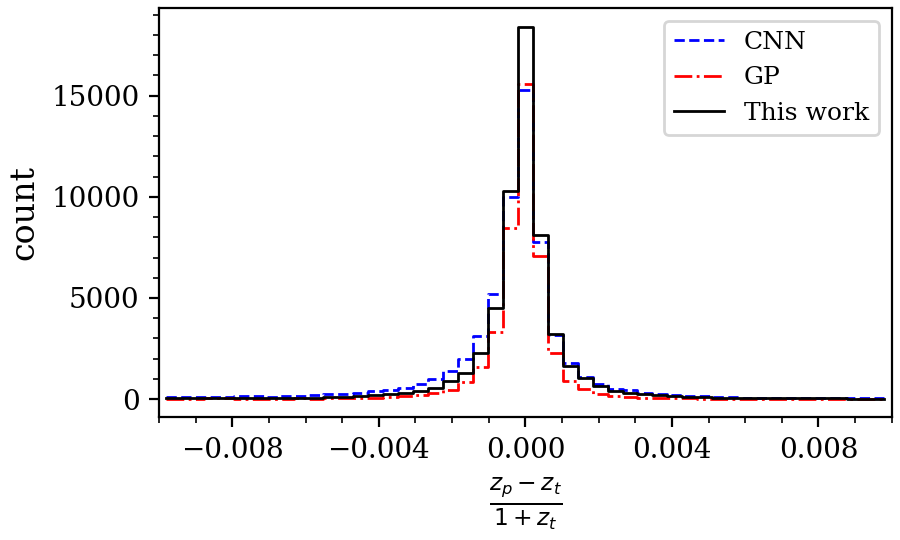}
    \includegraphics[width=\columnwidth]{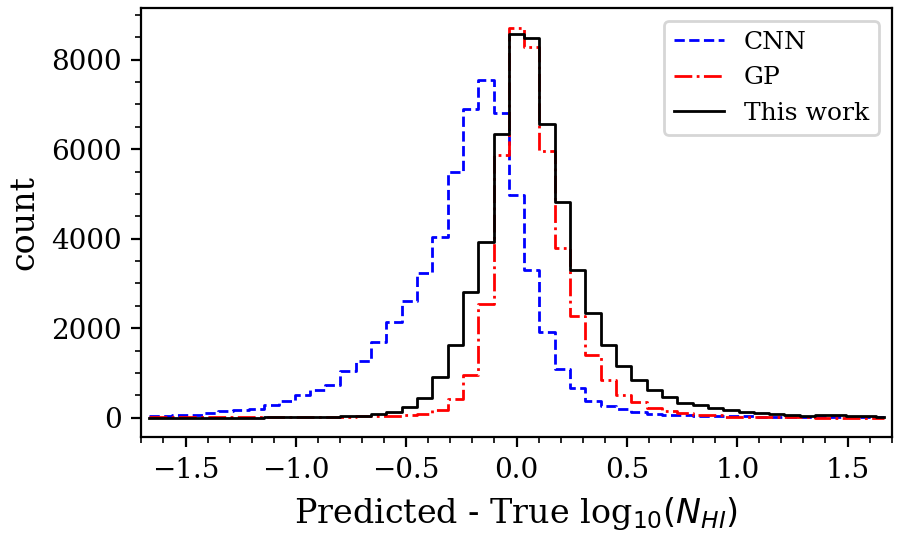}
    \caption{ Histograms illustrating the parameter accuracy of the DLA Toolkit, GP DLA finder, and CNN DLA finder for the true positive $\log_{10} (N_{\texttt{HI}})>20.3$ detections. The top panel shows the offset of the predicted $z$ from truth. All methods have an average offset consistent with zero. The bottom panel shows the difference between predicted and true $\log_{10}(N_\texttt{HI})$ values. The average offset is $0.100$ for the DLA Toolkit, $0.073$ for the GP DLA finder, and $-0.222$ for the CNN DLA finder.
        }
    \label{fig:z_nhi_hists}
\end{figure}

\begin{figure}[t]
\centering
    \includegraphics[width=\columnwidth]{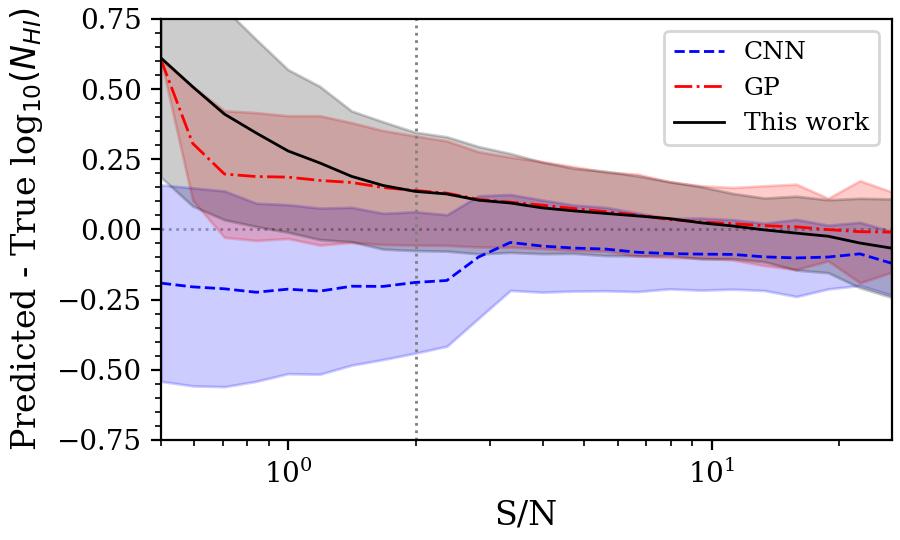}
    \caption{The average offset of predicted $N_\texttt{HI}$ from truth as a function of S/N for the DLA Toolkit, GP DLA finder, and CNN DLA finder for the true positive $\log_{10} (N_\texttt{HI})>20.3$ detections. The shaded regions are the standard deviation within each S/N bin. The vertical dashed line indicates S/N~$>2$. 
        }
    \label{fig:nhi_diff_vs_snr}
\end{figure}

As shown in Figure~\ref{fig:z_nhi_hists}, the majority of redshift estimates are accurate well within the boundary set by Equation~\eqref{eq:z_req}, with an average offset consistent with zero. 
Figure~\ref{fig:z_nhi_hists} also indicates that the DLA Toolkit slightly over-predicts column density on average by $\Delta \log_{10} (N_\texttt{HI}) = 0.100~(\sigma = 0.286)$. However, the $N_{\texttt{HI}}$ accuracy is S/N-dependent, and the average offset improves with increasing S/N as shown in Figure~\ref{fig:nhi_diff_vs_snr}. When restricting to S/N~$>2$, the average offset of column density from the truth value improves to $\Delta \log_{10} (N_\texttt{HI}) = 0.039~(\sigma = 0.207)$.

\begin{figure}[t]
\centering
	\includegraphics[width=\columnwidth]{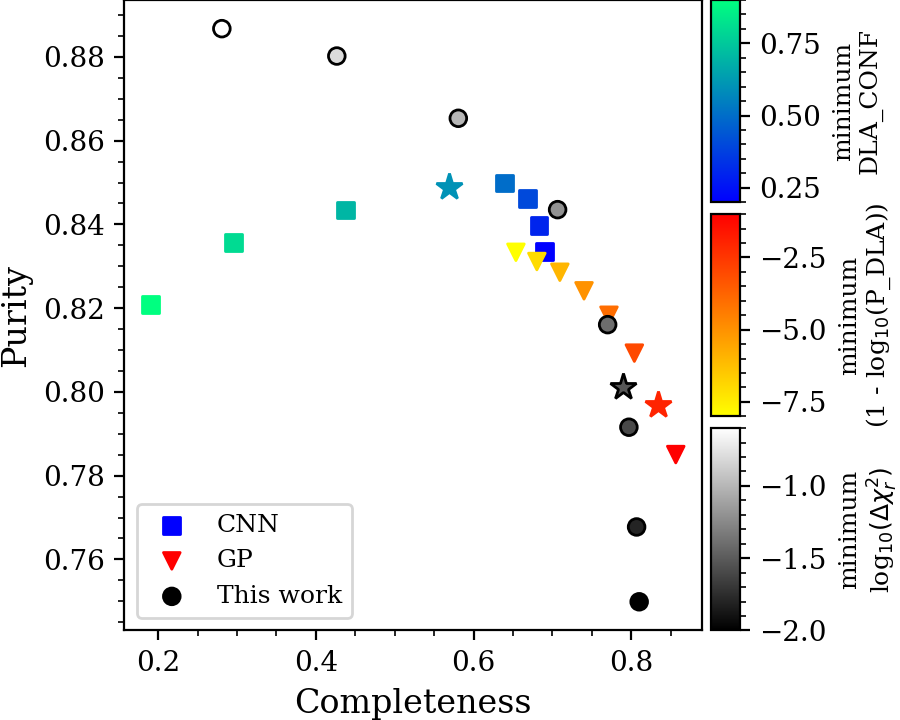}
    \caption{The purity and completeness of DLA samples for different cuts in detection confidence. The circle points correspond to the DLA catalog from the DLA Toolkit and are colored by the minimum $\Delta \chi_r^2$ to constitute a detection. The triangle (square) points correspond to the GP (CNN) DLA finder and are likewise colored by minimum \texttt{P\_DLA} (\texttt{DLA\_CONFIDENCE}) for detection. All DLA samples are limited to S/N~$>2$ and predicted $\log_{10} (N_\texttt{HI}) > 20.3$. The stars are the thresholds for the DESI DR2 BAO measurement. \\
        }
    \label{fig:purity_v_completeness}
\end{figure}

The purity of the DLA sample is defined as the number of true positives divided by the number of all detections. The completeness is the number of true positives divided by the total number of DLAs in the truth catalog. We first assess how purity and completeness depend on the $\Delta \chi_r^2$ threshold for detection (see Equation~\eqref{eq:chi2}). We compute these metrics after cutting the catalog on $\Delta \chi_r^2 > \Delta \chi_{r,min}^2$ for several values of $\Delta \chi_{r,min}^2$ spanning $0.01-0.25$. We only consider sightlines with a quasar S/N $>2$ since performance rapidly degrades below this. The S/N dependence is discussed in greater detail later in this section. As expected, the sample becomes more pure and less complete as the threshold $\Delta \chi_r^2$ increases, shown in Figure~\ref{fig:purity_v_completeness}. We elect a detection threshold $\Delta \chi_r^2 > 0.03$, indicated by the star, in all further analyses to balance the purity and completeness tradeoff. This choice yields a sample that is 80\% pure and 79\% complete.

\begin{figure}
\centering
    \includegraphics[width=\columnwidth]{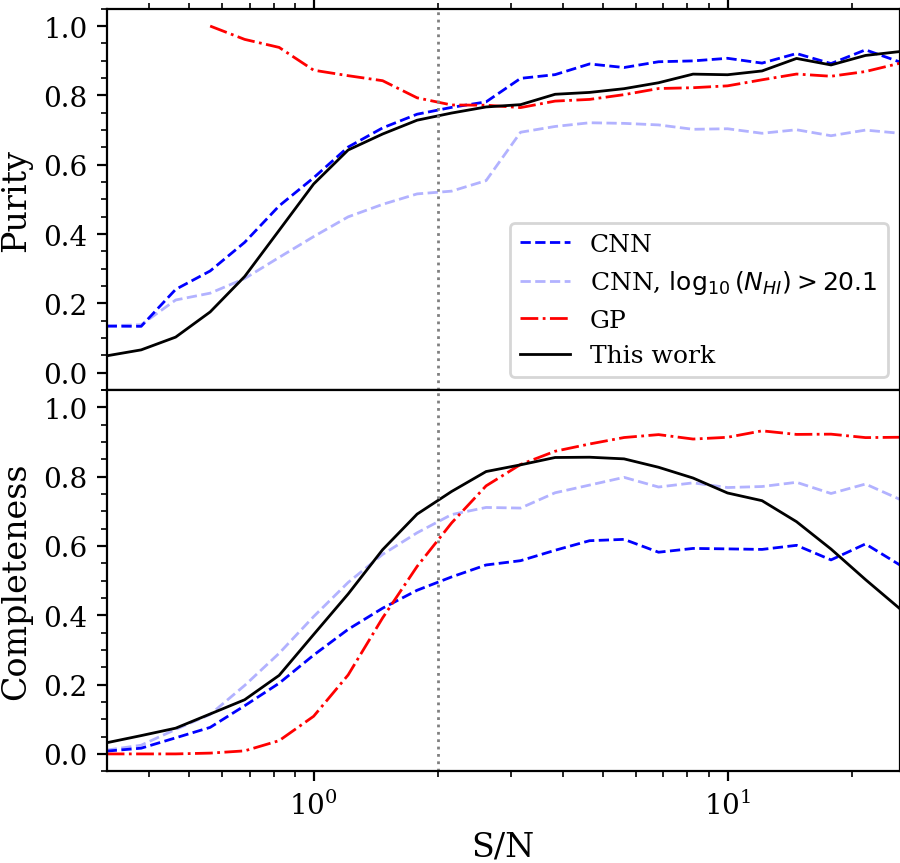}
    \caption{
        The purity and completeness of DLA samples as a function of S/N from the DLA Toolkit, GP DLA Finder, and CNN DLA finder. The samples are cut on confidence and predicted $\log_{10}(N_{\texttt{HI}}) > 20.3$, except the alternative CNN line as annotated. The vertical dashed line indicates S/N~$>2$.
        }
    \label{fig:snr_purity_completeness}
\end{figure}

We next evaluate the dependence of purity and completeness on S/N with the detection threshold of $\Delta \chi_r^2 > 0.03$. As evidenced in Figure~\ref{fig:snr_purity_completeness}, the purity increases rapidly up to S/N~$\approx 2$. The purity stabilizes beyond this point, increasing moderately from $\sim$75\% to $\sim$90\%. We observe a similar rapid increase in completeness at S/N~$<2$. In contrast to purity, completeness does not stabilize and actually begins to decrease around S/N~$\approx 5$, particularly for S/N~$\geq 10$. Approximately 22\% of the truth DLA sample are in S/N~$>5$ quasar spectra, where the average completeness of the DLA Toolkit is 77\%. Roughly 8\% of the truth DLAs are in quasar spectra with S/N~$>10$, where the DLA Toolkit has an estimated purity of 65\%. As such, the actual number of missed high S/N DLAs is relatively small compared to the full sample size. 

An analysis of the missed S/N~$>10$ DLAs reveals an excess of flagged fits. In high S/N spectra, the DLA Toolkit often finds better solutions (in a $\chi^2$ sense) from fitting two lower column density DLA profiles to a single true DLA trough instead of one DLA profile with a larger, more accurate column density. These fits are flagged owing to their poor parabola fits and thus discarded by our cuts. These dual solutions to single DLA troughs are driven by local minima, indicating the $\chi^2$ relaxation process is less robust at the highest S/N. We discuss this failure mode, its impact on Ly$\alpha$ clustering measurements, and steps for remediation in Section~\ref{sect:future}.

\begin{figure*}
\centering
    \includegraphics[width=\columnwidth]{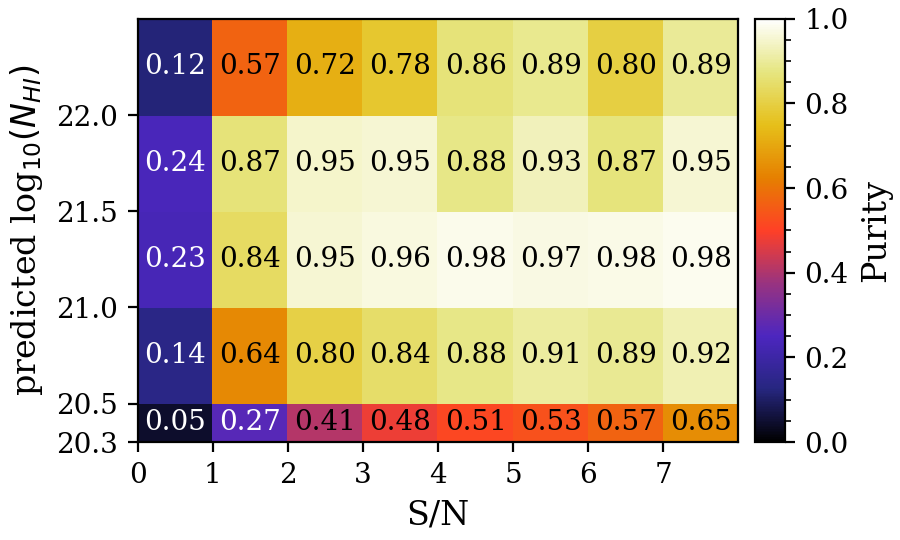}
    \includegraphics[width=\columnwidth]{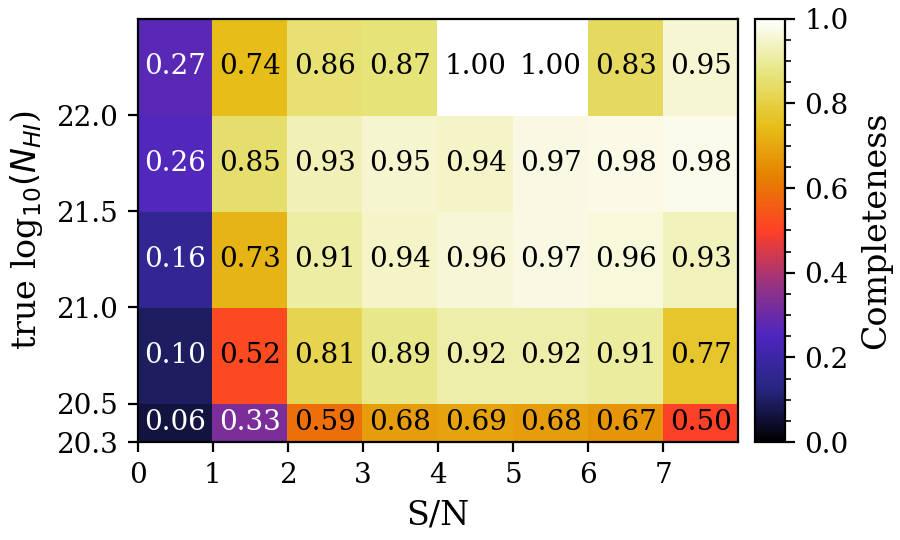}
    \caption{
        Purity (left) and completeness (right) values with the DLA Toolkit for different S/N and $\log_{10}(N_{\texttt{HI}})$ bins. 
        The $\log_{10} (N_\texttt{HI}) > 22.0$ bins contain $\leq 35$ objects each, with the $4<$~S/N~$<5$ and $5<$~S/N~$<6$ bins containing only 11 and 7 DLAs, respectively. 
        }
    \label{fig:tmp_matrix}
\end{figure*}

Our final test checks the dependence of purity and completeness on DLA column density, shown in Figure~\ref{fig:tmp_matrix}. Since $N_{\texttt{HI}}$ accuracy is S/N dependent, we check the purity and completeness in bins on both parameters. For S/N~$>2$ and $\log_{10} (N_\texttt{HI}) > 20.5$, purity nearly ubiquitously exceeds 80\% and completeness is generally above 85\%. As expected, we observe degraded performance at the lowest S/N and $N_{\texttt{HI}}$ values. 
The purity (completeness) is relatively low in the $20.3 < \log_{10} (N_\texttt{HI}) < 20.5$ bins owing to scatter in the predicted $\log_{10} (N_\texttt{HI})$ values causing sub-DLAs (DLAs) to fall above (below) the 20.3 minimum requirement. 

There is an apparent excess of false positive detections for high column densities ($\log_{10} (N_\texttt{HI}) > 22$). We visually inspected the model fit from the DLA Toolkit to several spectra in these bins. A large fraction of the false positives arises from the DLA Toolkit placing a DLA solution on the blue wing of the Ly$\alpha$ emission line to compensate for emission line diversity. This type of false positive could be removed by requiring the predicted DLA redshift to have a larger offset from $z_{QSO}$ than is currently imposed by the DLA Toolkit (see the definition of $z_{max}$ in Section~\ref{subsec:dlatoolkit}); however, doing so would lower the completeness at long wavelengths in Ly$\alpha$ forest thus increasing contamination from unmasked DLAs in the BAO analysis. For this reason, we maintain the current $z_{max}$ limit. A more robust solution would be a physicality check of DLA solutions at these wavelengths, such as confirming the observed flux is approximately zero inside the trough. Improvements to DLA detection accuracy in this regime will be explored in future work.

Lastly, when including BAL quasars in the test sample the DLA Toolkit achieves approximately the same completeness, but the purity decreases to 59.4\% for $\Delta \chi_r^2 > 0.03$ and S/N~$>2$. The trends for purity and completeness with S/N are relatively unchanged from the BAL-free sample, but the purity curve is shifted to lower values. The $N_\texttt{HI}$ and $z$ accuracy are comparable to the BAL-free sample. Table~\ref{tab:pc_meas} summarizes the purity and completeness metrics for both the full sample and the BAL-free sample.

The reduced purity compared to the nominal BAL-free test sample is likely due to imperfect masking. 
The BAL masking procedure (see Section~\ref{subsec:dlatoolkit}) assumes all BAL quasars exhibit broad absorption from C~\textsc{iv}  and that the C~\textsc{iv} BAL velocity profile is identical to the absorption profiles imprinted by other ions. The former is reasonable, as the presence of broad C~\textsc{iv} absorption typically defines the BAL quasar subtype \citep[e.g.,][]{hall02} in addition to being relatively easy to detect compared with other common BAL transitions. The latter assumption, however, perhaps oversimplifies complex dynamics of the BAL-producing structures \citep[e.g.][]{hamann19,masribas19a}. Furthermore, we use the C~\textsc{iv} BAL profile to generate masks only for the following transitions: C~\textsc{iv}, Si~\textsc{iv}, N~\textsc{v}, Ly$\alpha-\delta$, C~\textsc{iii} (977,1175), P~\textsc{v}, S~\textsc{iv}, O~\textsc{vi}, O~\textsc{i}, and N~\textsc{iii}. These lines are selected from spectral stacking analyses of C~\textsc{iv} outflows \citep[e.g.][]{masribas19b} and follow the convention of the Ly$\alpha$ forest BAO analysis. Other transitions (e.g., Si~\textsc{ii} and Al~\textsc{iii}) may be present but are unmasked, increasing the false positive potential and reducing the quality of the null fit. While the BAL masking evidently does not eliminate DLA/BAL confusion, the DLA Toolkit appears more robust against this type of contamination when compared to the GP DLA finder, which also relies on eigenspectra to construct a null model but does not mask for BALs (reflected in Table~\ref{tab:pc_meas}).

 \begin{table*}
     \centering
     \centerline{
     \begin{tabular}{|l|c|c|c|c|}
         \hline
                      & This Work              & GP                        & CNN                        & \multirow{2}{*}{Combined} \\
                      & $\Delta \chi_r^2 > 0.03$ & \texttt{P\_DLA}~$>0.99$   & \texttt{DLA\_CONF}~$>0.6$  &  \\
                      \hline
         Purity       & 80.1\%                 & 79.7\%                    & 84.9\%                     & 84.8\%  \\ \hline
         Purity (BAL) & 59.4\%                 & 53.6\%                    & 66.0\%                     & 66.4\%  \\ \hline
         Comp.        & 79.0\%                 & 83.4\%                    & 56.9\%                     & 79.0\%  \\ \hline
         Comp. (BAL)  & 77.3\%                 & 83.4\%                    & 55.9\%                     & 77.6\%  \\ \hline
     \end{tabular}
         }
     \caption{
         The purity and completeness of DLA samples from the DLA Toolkit, GP DLA Finder, CNN DLA Finder, and the combination of the three methods (as outlined in Section~\ref{BAO_combined_catalog}) for the \texttt{Ly$\alpha$CoLoRe} mock spectra sample. All DLA samples are from quasar spectra with S/N~$>2$ and have predicted $\log_{10} (N_\texttt{HI}) > 20.3$. 
         }
     \label{tab:pc_meas}
 \end{table*}

\subsection{Comparison to other DLA algorithms}
\label{subsect:comparison_of_techs}

The GP DLA finder from \citet{ho21} and the CNN DLA finder from \citet{wang22} were used in conjunction to identify DLA contamination in the DESI DR1 Ly$\alpha$ forest BAO measurement \citep{DESI-Y1KP6}. The current version of the GP model was trained on quasar spectra from the extended Baryon Acoustic Oscillation Survey \citep[eBOSS;][]{eboss}.\footnote{The GP model used in this work is a Python-translated version of the original MATLAB GP code, with optimizations to improve speed.  The software is available at \url{https://github.com/jibanCat/desi_gpy_dla_detection}} The current CNN version was trained on simulated quasar spectra mimicking the first year of observations with the DESI survey. We run these DLA finders on the same simulated quasar sample that was used to evaluate the DLA Toolkit in Section~\ref{subsect:mock_validation} and compare their performances. We refer the reader to \cite{ho21, wang22} for complete details on these two methods.

To ensure analogous results from each technique, we perform a series of cuts on the output DLA catalogs from the GP and CNN DLA finders on the \texttt{Ly$\alpha$CoLoRe} mocks. We remove detections with predicted $\log_{10} (N_\texttt{HI}) < 20.3$ and restrict the DLA redshift range to that defined by Equation~\eqref{eq:zmin} and Equation~\eqref{eq:zmax}. We then remove any detections flagged as problematic by the GP. The CNN does not maintain any flags, so we retain all detections from this algorithm that pass the redshift and column density cuts. As with the DLA Toolkit, we consider a GP or CNN DLA detection a true positive if the predicted redshift uniquely satisfies Equation~\eqref{eq:z_req} for any DLA in the $\log_{10} (N_\texttt{HI}) > 20.3$ 
truth catalog. 

Figure~\ref{fig:z_nhi_hists} illustrates that, on average, the GP and CNN methods achieve equally precise redshift estimates as the DLA Toolkit. The average offset of the predicted GP and CNN redshifts from truth is consistent with zero, similar to what was reported by \cite{wang22}.
The GP, however, produces a tighter distribution for the offset of predicted $N_{\texttt{HI}}$ values from truth than either the CNN or the DLA Toolkit. The average offset of GP predicted column density from truth is $\Delta \log_{10} (N_\texttt{HI}) = 0.073~(\sigma=0.198)$. The CNN, in contrast, provides the least reliable column density predictions of the three methods, on average, with a mean offset of $\Delta \log_{10} (N_\texttt{HI}) = -0.222~(\sigma=0.475)$ \citep[][previously reported a bias on column density]{parks18,chabanier22}. 

We next compare the dependence of $N_{\texttt{HI}}$ accuracy on S/N for the three methods in Figure~\ref{fig:nhi_diff_vs_snr}. The DLA Toolkit and GP tend to overestimate $N_\texttt{HI}$ in a relatively similar fashion that improves with increasing S/N. This may be a consequence of the difficulty in modeling intrinsic quasar continua at lower S/N, which both approaches rely on when estimating column density. Meanwhile, the CNN tends to underestimate $N_\texttt{HI}$ by a consistent average magnitude of $\Delta \log_{10} N_\texttt{HI} \approx 0.08$ at S/N~$>3$ and $\Delta \log_{10} N_\texttt{HI} \approx 0.205$ at S/N~$<2$. It is clear that $N_\texttt{HI}$ accuracy is degraded for all methods at low S/N. Applying a lower limit S/N~$>2$, the average offset of column density from the truth value improves to $\Delta \log_{10} (N_\texttt{HI}) = 0.057~(\sigma = 0.188)$ for the GP and $\Delta \log_{10} (N_\texttt{HI}) = -0.164~(\sigma = 0.232)$ for the CNN.  Figure~\ref{fig:nhi_diff_vs_snr} also makes it apparent that the improved average $N_\texttt{HI}$ accuracy of the GP over the DLA Toolkit arises in the S/N extremes (S/N~$\lesssim1.5$ and S/N~$\gtrsim18$). 

The GP and CNN DLA finders both have an output variable that quantifies detection significance for candidate DLAs, similar to the DLA Toolkit's $\Delta \chi_r^2$. The GP reports a \texttt{P\_DLA} value, and the CNN reports a \texttt{DLA\_CONFIDENCE} value for each detection. Both parameters are on a scale of zero to one where larger values correspond to higher confidence. \texttt{P\_DLA} exhibits a rather bimodal distribution, returning values greater than 0.9 or equal to zero. The latter is equivalent to a non-detection. For the CNN, the distribution of detections generally decreases with increasing \texttt{DLA\_CONFIDENCE} value, aside from a spike of detections with \texttt{DLA\_CONFIDENCE}~$=1.0$.
 
Figure~\ref{fig:purity_v_completeness} demonstrates how the purity and completeness of the GP and CNN DLA samples depend on their respective detection thresholds. The behavior of the CNN as the threshold \texttt{DLA\_CONFIDENCE} increases suggests that \texttt{DLA\_CONFIDENCE} is not necessarily a good indicator of DLA probability. We define a DLA detection by \texttt{P\_DLA}~$>0.99$ for the GP and \texttt{DLA\_CONFIDENCE}~$>0.6$ for the CNN. Using these thresholds, the GP provides a DLA sample that is 79.7\% pure and 83.4\% complete, and the CNN provides a DLA sample that is 84.9\% pure and 56.9\% complete.

\begin{figure*}[t]
\centering
	\includegraphics[width=2\columnwidth]{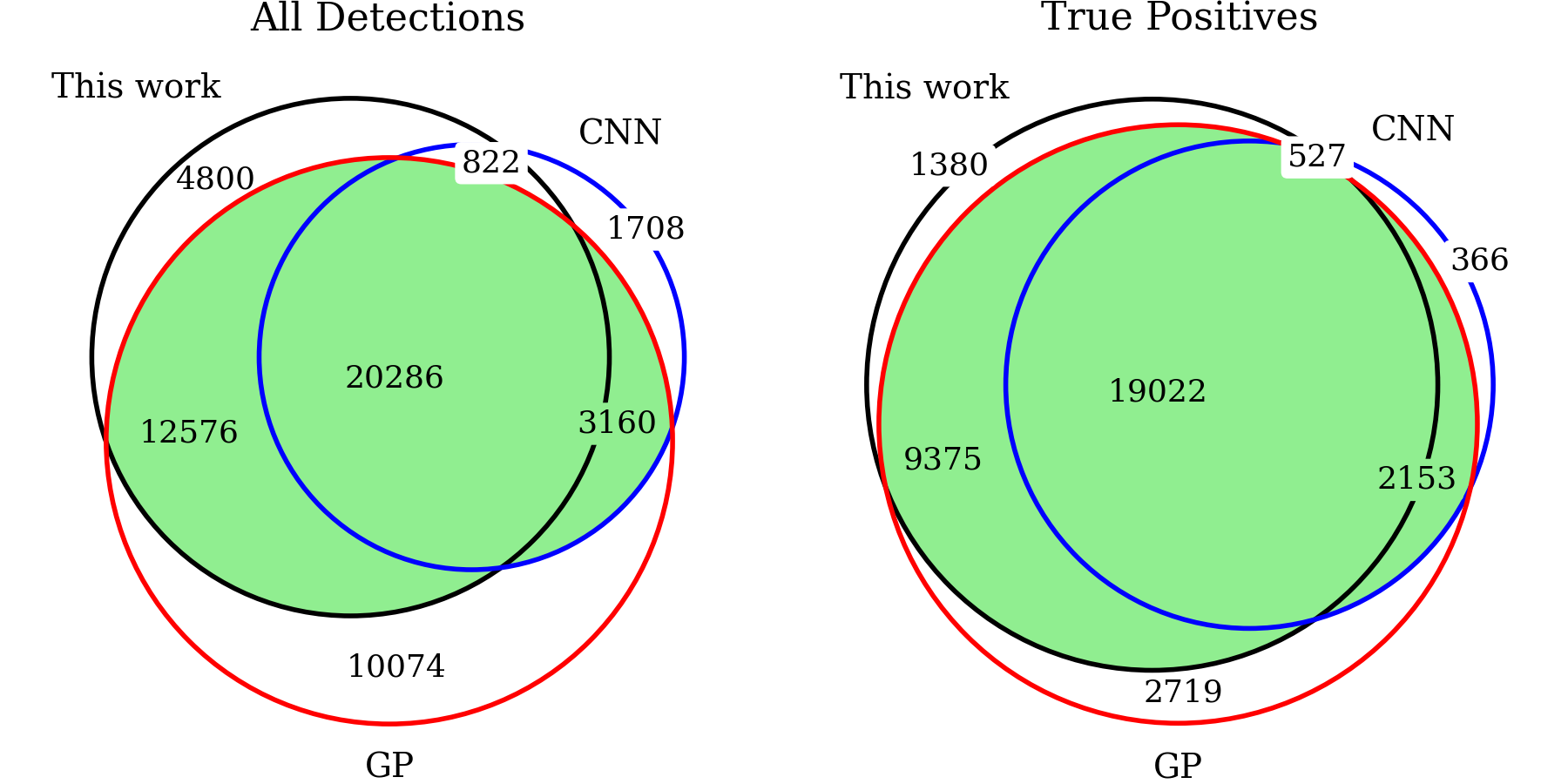}
    \caption{
        Venn diagram of all detections (left) and the true positive detections (right) by the DLA Toolkit, GP, and CNN methods where S/N $> 2$ and BAL sightlines are excluded. 
        These detections are required to meet $\Delta \chi_r^2 > 0.03$, \texttt{P\_DLA} $> 0.99$, and/or \texttt{DLA\_CONFIDENCE} $> 0.6$ depending on method. 
        We also require $\log_{10}(N_{\texttt{HI}}) > 20.3$.
        The green shaded region represents the detections selected by Equation~\eqref{eq:combined_cat}.
    }
    \label{fig:venn_mock}
\end{figure*}

The top panel of Figure~\ref{fig:snr_purity_completeness} shows how the purity of these samples depends on S/N with the chosen detection thresholds. The GP purity is fairly consistent at $\sim$80\% for S/N~$>2$, but reports no detections below S/N~$<0.5$. The lack of low-S/N detections may be driven by inadequate training of the GP model on noisy data. Since the GP relies on a Bayesian framework, its predictions become prior-driven when the model lacks sensitivity to certain data regimes, i.e., favoring non-DLA detection by default. The CNN exhibits a similar trend in purity to the DLA Toolkit but is up to $\sim$10\% more pure at the S/N~$<1$ and roughly $5\%$ more pure between $3 < \text{S/N} < 10$. DLA samples from the CNN are also about 5\% more pure than the GP samples for S/N~$>2$.

The trends of completeness with S/N are shown in the bottom panel of Figure~\ref{fig:snr_purity_completeness}. The GP achieves roughly uniform completeness of $\sim$90\% for S/N~$>4$. The completeness of the CNN levels out earlier at S/N~$\approx2$, hovering around 55\%. The DLA Toolkit outperforms the GP and CNN at S/N~$<3$, after which the GP provides the most complete DLA sample. A summary of the purity and completeness metrics for the GP and CNN DLA finders is given Table~\ref{tab:pc_meas}. 

Table~\ref{tab:pc_meas} also includes the performance of the GP and CNN DLA finders on the quasar sample that includes BAL sightlines. As with the DLA Toolkit, the completeness with the GP and CNN DLA finders is relatively unaffected by BAL sightlines. However, as evidenced by the purity values, all algorithms suffer from DLA/BAL confusion. The CNN is the most robust of the three against false positives caused by BALs. This is likely because it does not rely on eigenspectra to construct an intrinsic model of the quasars, unlike the GP and DLA Toolkit. These models cannot capture the spectral variation caused by BALs and, without masking or with insufficient masking, tend to produce poor reconstructions.

The low completeness of the CNN reported in this work is likely driven by its systematic underestimation of $N_\texttt{HI}$. We show in Figure~\ref{fig:snr_purity_completeness} that relaxing the column density threshold to predicted $\log_{10} (N_\texttt{HI}) > 20.1$ can boost the CNN's completeness but at the cost of lower purity. With this lower $N_{\texttt{HI}}$ minimum, the purity (completeness) decreases (increases) $\sim 20\%$ above S/N$> 2$. Since the focus of this work is identifying DLA contamination for BAO, we choose to keep the $\log_{10} (N_\texttt{HI}) > 20.3$ cut-off to minimize the loss of uncontaminated forest regions. Further, we expect the GP and DLA Toolkit will compensate for CNN in this regime (see Section~\ref{BAO_combined_catalog}). We refer the reader to \cite{parks18} and \cite{chabanier22} for discussions on de-biasing $N_\texttt{HI}$ from the CNN, but caution that de-biasing procedures should be updated to reflect the new training \citep{wang22}. We leave more sophisticated work on de-biasing the CNN $N_{\texttt{HI}}$ and subsequent effects on DLA detections to future works. 

Differences in our reported performance metrics compared to \citet{wang22}, who evaluated both the GP and CNN methods on DESI DR1 mock spectra, are likely due to a combination of factors.  First, the DESI DR2 mocks used in this study feature improved realism relative to the mock spectra used in their analysis \citep[see][for details on the mock improvements]{casas_mocks}. We also do not consider sub-DLA column densities ($\log_{10} N_\texttt{HI} < 20.3$) in our assessment, whereas the DLA samples in \cite{wang22} are defined by $\log_{10} N_\texttt{HI} > 20.0$. The lower column density threshold results in a substantially larger sample size than with the canonical DLA column density cut-off used in this work. We additionally find larger offsets of predicted $N_\texttt{HI}$ from truth than was reported by \cite{wang22}, particularly for the CNN method, even when restricting to S/N~$>3$. Finally, we use a lower S/N cut-off and higher detection significance thresholds for both methods for our final catalog, informed by the results of our analyses.

Figure~\ref{fig:venn_mock} explores the overlap between the DLA catalogs produced by the DLA Toolkit, GP, and CNN methods for  $\log_{10}(N_{\texttt{HI}}) > 20.3$ detections. Motivated by the previous tests, we require $\Delta \chi_r^2 > 0.03$, \texttt{P\_DLA} $> 0.99$, \texttt{DLA\_CONFIDENCE} $> 0.6$, and S/N~$>2$. 
We associate detections between catalogs if they have predicted $z$ values within 800 km~s$^{-1}$, following the convention of \citep[][]{DESI-Y1KP6}.\footnote{This $\Delta z$ tolerance works well for most column densities; However, at the highest column densities, we show that a relaxed tolerance for associating detections between catalogs provides better performance in Appendix~\ref{appx:diff_combined_strategy}.}

The distribution of candidate DLAs found by at least one of the DLA Toolkit, GP, or CNN methods is shown on the left in Figure~\ref{fig:venn_mock} while the right shows the subset which are true DLAs. The largest subset of candidate DLAs shown in Figure~\ref{fig:venn_mock} corresponds to those detected by all three methods. This subset has a high purity of 93.8\%, demonstrating how combining techniques helps to filter out contamination from the individual catalogs. 
The next largest subset is candidate DLAs detected by the DLA Toolkit and the GP method with a purity of 74.5\%. 
After this, the next largest subset is candidate DLAs detected by only the GP method with a purity of 27.0\%. 
The two subsets corresponding to shared GP and CNN candidate DLAs or shared DLA Toolkit and CNN candidate DLAs are relatively smaller but have 68.1\% and 64.1\% purity, respectively. 
These results suggest that while detections made by all three methods are the most pure, each method contributes unique identifications. Furthermore, the catalogs can be used as cross-checks for each other to remove a large number of false positives. This highlights the importance of using multiple approaches for a more comprehensive DLA catalog.

\section{DLA Catalogs from the DLA Toolkit}
\label{sect:catalogs}
We run the DLA Toolkit on the quasar samples from DESI DR1 and DR2 described in Section~\ref{subsec:desi_data}, restricting both samples to $2.0<z_{QSO}<4.25$. Flagged detections are removed from the output catalogs and therefore not considered in the statistics reported in the following subsections. We refer the reader to the performance metrics reported in Section~\ref{subsect:mock_validation} as a guideline for selecting and using DLA samples from these catalogs. Since simulated spectra cannot capture the full diversity observed in real quasar spectra, the purity and completeness measurements should be treated as approximate upper bounds. Candidate DLAs can be matched via \texttt{TARGETID} to the corresponding quasar catalog to remove BAL sightlines and enhance the sample purity. 

\subsection{The DR1 Catalog}
\label{subsec:dr1_subset_catalog}

After applying the quasar redshift cut, the DR1 sample consists of 520,745 sightlines. The DLA Toolkit records 74,918 detections for this sample, of which 60,565 have predicted $\log_{10} N_{\texttt{HI}}>20.3$. Approximately 1\% of quasar sightlines contain more than one candidate DLA. 

\begin{figure}[t]
\centering
	\includegraphics[width=\columnwidth]{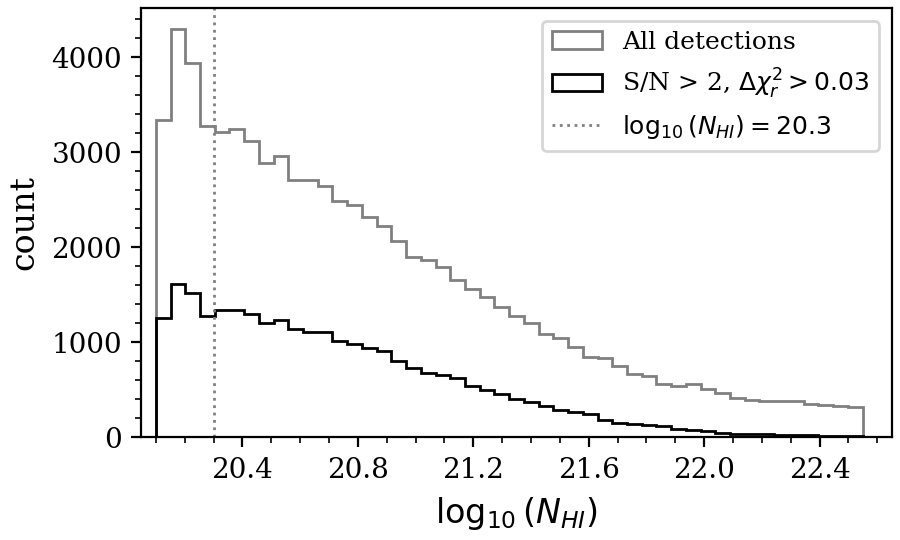}
    \includegraphics[width=\columnwidth]{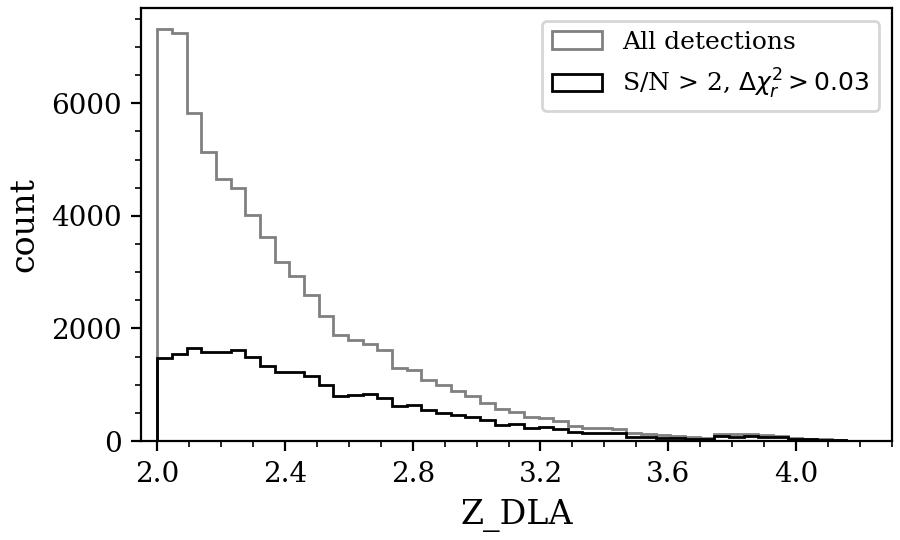}
    \caption{The predicted redshift and $\log_{10} (N_\texttt{HI})$ distribution of the DLA Toolkit catalog for DESI DR1. The vertical line is the DLA column density definition.
        }
    \label{fig:dr1_distributions}
\end{figure}

The best performance is expected for S/N~$>2$, as evidenced throughout Section~\ref{subsect:mock_validation}. Using this S/N limit and a $\Delta \chi_r^2 > 0.03$ threshold, 26,651  DLA candidates remain in the catalog with 21,173 having predicted $\log_{10} N_{\texttt{HI}}>20.3$ (from 235,856 S/N~$>2$ sightlines). Figure~\ref{fig:dr1_distributions} shows the distribution of predicted column density and redshift for the full sample and the sample optimized with S/N and $\Delta \chi_r^2$ cuts.

The DR1 DLA catalog from the DLA Toolkit, along with the corresponding quasar catalog, is available concurrently with the full DESI DR1 \citep{DESI_DR1}. The DLA catalog contains the information summarized in Table~\ref{tab:cat_columns}. The catalog complements the concordance DLA catalog presented in the forthcoming paper by \cite{dr1_dlas}, which includes DLAs found with the CNN and GP methods. 

\subsection{The DR2 Catalog}
\label{subsect:dr2_catalog}

The DESI DR2 sample contains 942,946 quasar sightlines that satisfy $2.0<z_{QSO}<4.25$. The DLA Toolkit returns 137,222 candidate DLAs with predicted $\log_{10} (N_\texttt{HI})>20.3$. Restricting to the S/N~$>2$ (455,360 quasar sightlines) leaves 66,955 candidate DLAs with predicted $\log_{10} (N_\texttt{HI})>20.3$ remaining in the sample. Further requiring $\Delta \chi_r^2 > 0.03$ results in an optimized sample of 52,575 candidate DLAs. Similar to the DR1 catalog, roughly 1\% of sightlines have more than one candidate DLA.

\section{The DR2 Combined Catalog for Lyman-Alpha Forest BAO}
\label{BAO_combined_catalog}
This section presents the construction of the DLA catalog for the DR2 Ly$\alpha$ forest BAO measurement \citep{DESI_lyaBAO}. The wings of DLA absorption profiles, which can extend for thousands of km~s$^{-1}$, not only compromise the ability to extract the neutral hydrogen density field but increase noise in the correlation function and alter its broadband shape \citep[e.g.][]{fontribera12a}. DLAs also cluster more strongly than the Ly$\alpha$ forest \citep[e.g.][]{fontribera12b,perez18a,perez23}, increasing the bias of the correlation function. It is essential to efficiently identify DLAs for the BAO analysis, so they may be masked and their impact on the correlation function mitigated. We combine output DLA catalogs from the GP, CNN, and DLA Toolkit on the DESI DR2 quasar sample to construct a highly complete and pure catalog. We aim to optimally balance these two metrics as to reduce contamination without erroneous loss of forest pixels.
 
We combine output catalogs from the CNN and GP DLA finders on the DR2 quasar sample with the DLA Toolkit catalog from Section~\ref{subsect:dr2_catalog} as follows. To begin, we match candidate DLAs between catalogs within 800 km~s$^{-1}$. We then remove detections from each method that do not meet the $\log_{10} (N_\texttt{HI}) > 20.3$ or confidence minimums. The chosen confidence minimums per method are informed by the results presented in Section~\ref{sect:performance}. We also require S/N~$>2$ since we expect degraded performance from all three algorithms at low S/N. In summary, detections ($det_i$) from each algorithm are defined following Equations~\eqref{eq:dlatoolkit_detection}, \eqref{eq:gp_detection}, and \eqref{eq:cnn_detection}. The number of candidate DLAs satisfying the detection criterion for each method is given in Table~\ref{tab:dr2_candidate_dlas}.

\begin{equation}
\label{eq:dlatoolkit_detection}
    \begin{aligned}
        det_{\texttt{Toolkit}} = 
        & (\log_{10}(N_{\texttt{HI-Toolkit}}) > 20.3) \\
        & \cap (\Delta \chi_r^2 > 0.03) \cap (\text{S/N} > 2)
    \end{aligned}
\end{equation}

\begin{equation}
\label{eq:gp_detection}
    \begin{aligned}
        det_{\texttt{GP}} = 
        & (\log_{10}(N_{\texttt{HI-GP}}) > 20.3) \\
        & \cap (\texttt{P\_DLA} > 0.99) \cap (\text{S/N} > 2)
    \end{aligned}
\end{equation}

\begin{equation}
\label{eq:cnn_detection}
    \begin{aligned}
        det_{\texttt{CNN}} = 
        & (\log_{10}(N_{\texttt{HI-CNN}}) > 20.3) \\
        & \cap (\texttt{DLA\_CONFIDENCE} > 0.6) \cap (\text{S/N} > 2)
    \end{aligned}
\end{equation}

Based on the catalog overlap discussion in Section~\ref{subsect:comparison_of_techs}, the final combined catalog requires a detection by both the GP and either one of the DLA Toolkit or the CNN. This decision aims to maximize completeness without sacrificing purity. The GP detection then sets the final predicted redshift and column density so that all candidate DLAs use a common z and $N_\texttt{HI}$ estimator. Equation~\eqref{eq:combined_cat} summarizes the logic for constructing the combined catalog. The final catalog has 41,152 candidate DLAs. We note that this catalog does not remove BAL sightlines.

\begin{equation}
    \begin{aligned}
        det_{\texttt{Combined}} = 
        det_{\texttt{GP}} \cap  (det_{\texttt{Toolkit}} \cup det_{\texttt{CNN}})
    \end{aligned}
    \label{eq:combined_cat}
\end{equation}

\begin{table}
     \centering
     \centerline{
     \begin{tabular}{|l|c|c|}
         \hline
& Total Candidate DLAs & non-BAL sightlines \\
\hline
DLA Toolkit &  52,575 & 35,939 \\
GP & 69,995 & 32,100 \\
CNN & 52,072 & 25,457 \\
\hline
Combined & 41,152 & 25,568 \\ 
\hline
         
     \end{tabular}
         }
     \caption{
         Number of candidate DLAs found in the DESI DR2 quasar sample by each method following the definitions in Equations~\eqref{eq:dlatoolkit_detection}$-$\eqref{eq:combined_cat}.}
     \label{tab:dr2_candidate_dlas}
 \end{table}

 \begin{figure*}
\centering
	\includegraphics[width=\columnwidth]{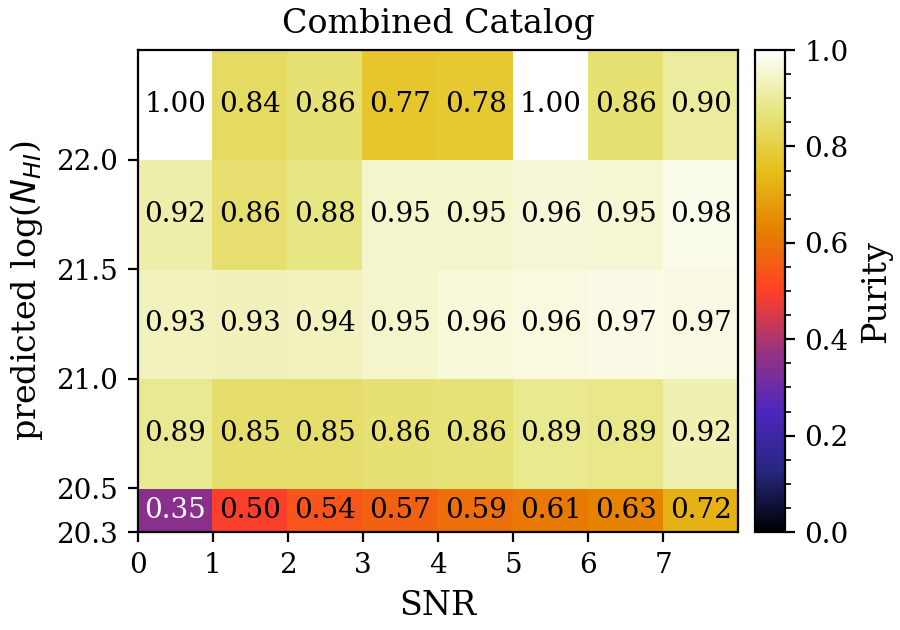}
    \includegraphics[width=\columnwidth]{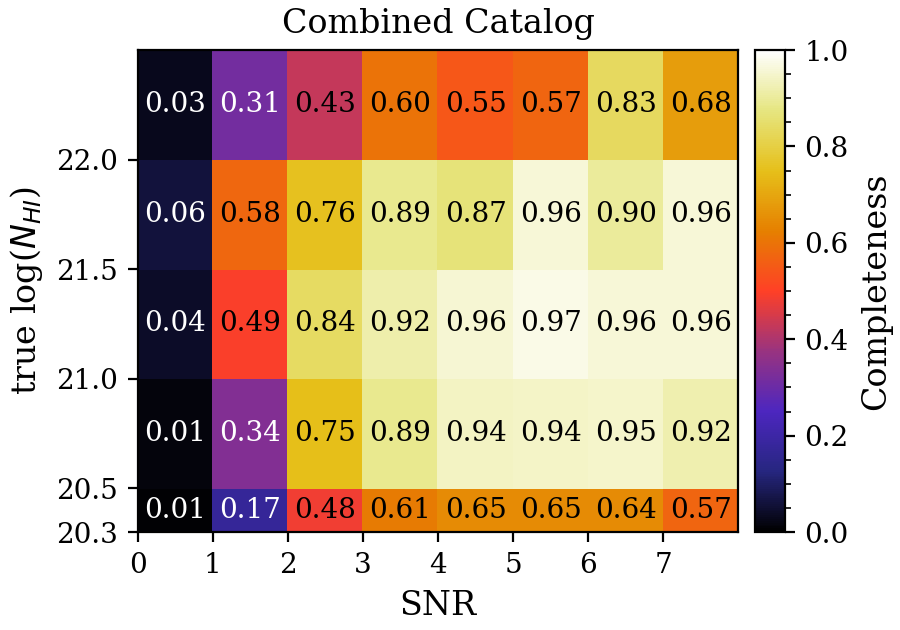}
    \caption{
        The purity (left) and completeness (right) of the combined DLA catalog for the DESI DR2 Ly$\alpha$ forest BAO measurement constructed following Equation~\eqref{eq:combined_cat} with the individual catalogs from the DLA Toolkit, GP, and CNN methods from Section~\ref{sect:performance}. An alternative strategy that improves completeness for large $N_{\texttt{HI}}$ values is presented in Appendix \ref{appx:diff_combined_strategy}, which may be more optimal for non-BAO use cases.
        }
    \label{fig:combined_matrix}
\end{figure*}

To assess the expected purity and completeness of the combined catalog, we apply Equation~\eqref{eq:combined_cat} to the mock DLA catalogs presented in Section~\ref{sect:performance} from the DLA Toolkit, GP, and CNN methods.  For this analysis, we exclude BAL sightlines since that type of contamination is removed separately by \cite{DESI_lyaBAO} in the BAO measurement. DLA detections satisfying Equation~\eqref{eq:combined_cat} are shown as the green shaded area in Figure~\ref{fig:venn_mock}. It has a purity of 84.8\% and a completeness of 79.0\%. These metrics are included in Table~\ref{tab:pc_meas} for comparison to the catalogs from the individual methods.
We validate our mock estimated purity with data stacks in Section~\ref{subsubsect:purity_validation}.

We considered an alternative combination strategy that requires detection by any two of the methods. This choice slightly increases (decreases) the completeness (purity) of the sample, as illustrated in Figure \ref{fig:venn_mock}; However, requiring a common method across all candidate DLAs in the combined catalog allows us to standardize the parameter predictions. The GP is the best choice for the common method owing to its greater $N_{\texttt{HI}}$ accuracy relative to the other two methods and its high completeness.

We also evaluated the DR1 BAO strategy which required $\texttt{CNN\_CONFIDENCE} > 0.5$ and $\texttt{P\_DLA} > 0.5$ for building a combined DLA catalog for DR2.
With these thresholds, we get 90.8\% purity and 61.0\% completeness at S/N~$>2$.
The accuracy of this catalog's redshift and column density is identical to our presented strategy, as both utilize the GP solutions for the final predictions. 
Thus, this new DR2 strategy increases completeness by roughly 18\% while only losing about 5\% in purity.

As a final analysis of the combined catalog using mock spectra, we evaluate purity and completeness in the same S/N and $\log_{10 }(N_\texttt{HI})$ bins as in Figure~\ref{fig:tmp_matrix}
for the DLA Toolkit.
These results are shown in Figure~\ref{fig:combined_matrix}.
The relatively low completeness at high $N_{\texttt{HI}}$ 
is due to a combination of factors: a known issue where the DLA Toolkit (and GP method) can fit two lower-$N_{\texttt{HI}}$ DLA profiles instead of one higher-$N_{\texttt{HI}}$ DLA profile to a single trough (discussed in Section~\ref{subsect:mock_validation} and Section~\ref{sect:future}), and the strict requirement that matched DLAs between catalogs must fall within 800~km~s$^{-1}$.
Relaxing this criterion to 3,000~km~s$^{-1}$, as done for determining true positives with Equation \eqref{eq:detection_def}, substantially improves completeness at high $N_{\texttt{HI}}$. Appendix~\ref{appx:diff_combined_strategy} presents the impact of this alternative matching criterion in more detail.

\subsection{Purity Validation with Spectral Stacking}
\label{subsubsect:purity_validation}
Spectroscopic stacking is a common tool for the study of well-understood absorber samples. \citet{Pieri2014} and \citet{Frank18} also showed that one can stack mixed samples of apparent absorption and, using the basic rules of atomic physics and spectroscopy, learn the mix of ionization species and noise that gave rise to the sample. In those articles lines were stacked to determine whether they were caused by metal doublets via the deviation from the normal 2:1 line ratio. In effect, one can test the purity with which the apparent lines found are any desired metal doublet.

 \begin{figure*}[t]
\centering
	\includegraphics[width=\linewidth]{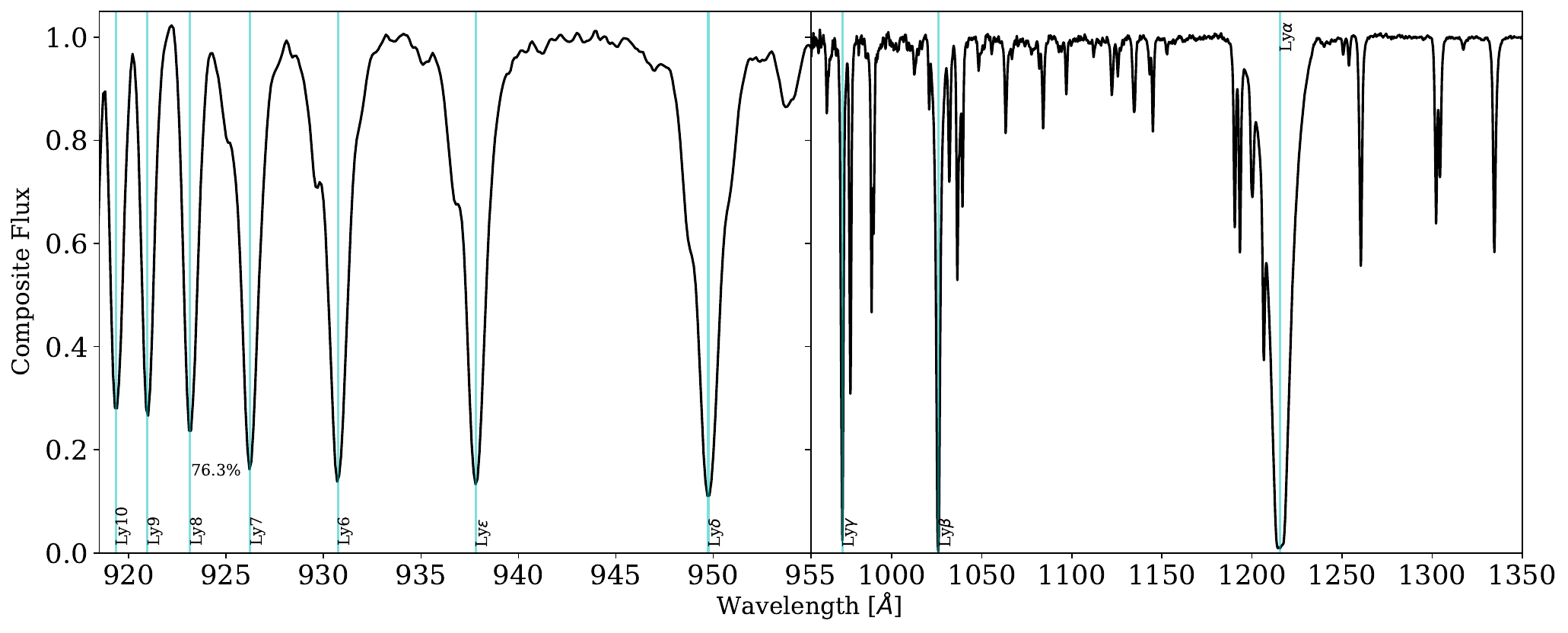}
    \caption{Composite spectrum of DLAs in the DR2 Combined Catalog, according to the properties defined in Section~\ref{subsubsect:purity_validation}. Lyman series lines are marked with vertical cyan lines. The percentage close to the Lyman-8 line represents the purity in the selected DLA sample for predicted $z\gtrsim2.9$.
        }
    \label{fig:CompositeCombinedCatalogue}
\end{figure*}

Here we use the Lyman series in place of metal doublets. We use the fact that all DLAs with a column density $\log_{10} (N_\texttt{HI}) > 20.3$ should show no transmission at the line center for at least the first 10 Lyman series lines \citep{dodorico01,omeara01}, even at DESI resolution. By construction, any Lyman series lines included in a DLA identification process will show no (or nearly no) transmission, but any Lyman series lines not included are available for this test of purity. However, a subtlety remains that must be accounted for; the typical contaminants of DLA samples are large complexes of strong Ly$\alpha$ forest absorption (along with noise). The Lyman series absorption associated with these complexes depends on the mix of column densities for each interloper. These complexes may show strong absorption at Ly$\alpha$ but as one moves up the Lyman series they converge to an expectation value of 100\% transmission. In a stack with a mix of real DLAs and these  interlopers,
the mean transmission at line center must increase as one moves up the Lyman series to asymptote to a constant value corresponding to a mixture of DLAs with 0\% expected transmission  
and interlopers with 100\% expected transmission. This procedure has been tested with mock spectra and appears to be unaffected by DLA redshift errors for all classifiers discussed here.
 
In the work that follows, this asymptote appears to occur at Lyman-8, such that Lyman-9 and Lyman-10 generate consistent results within the limits of the signal-to-noise of the stacked spectrum. Hence the purity of the stacked DLA sample is given by the measured flux decrement ($1-F$) divided by the expected flux decrement at Lyman-8 (unity).

The stacking procedure goes as follows: firstly, we choose which DLAs to stack. In this case, we chose absorbers in the DR2 combined catalog defined by Equation~\eqref{eq:combined_cat}, removing quasar sightlines flagged for BAL features. We require the DLA to lie between 911~\AA\ and 1205~\AA\ in the quasar's rest frame. The Lyman-8 absorption line redshifts into the DESI spectrograph wavelength coverage ($\lambda_{min}=3600$~\AA) for candidate DLAs with predicted $z\gtrsim 2.9$. As such, we only measure the purity of the subset with predicted $z \gtrsim 2.9$, corresponding to $\sim$22\% of the combined catalog after BAL sightlines were excluded. As a result, our test includes relatively few DLAs in the Ly$\beta$ forest and is a somewhat conservative estimation of DLA purity in the entire sample (since the forest opacity is higher at higher redshift and so DLA interlopers are more common).

After defining our sample for stacking, we choose a grid upon which to interpolate the data. In this case, we chose a grid spanning 910~\AA\ to 1350~\AA\ (to capture the Lyman Limit), with a periodicity of 0.25~\AA. This interval approximately matches the 0.8~\AA\ spacing of the DESI spectrographs redshifted to the DLA rest frame. 

Next comes the stacking itself. For each DLA, we shift the wavelength solution of its corresponding spectrum to the absorber rest frame, which is set by the predicted redshift from the GP for the combined catalog. The spectrum is then continuum normalized, and the flux and inverse variance are interpolated onto the stacking grid we defined beforehand. The mean value of the flux and inverse variance are calculated, giving rise to the stack. We perform a pseudo-continuum fit on the stack, allowing us to correct for systematic errors in the continuum fits as well as remove undesirable features from uncorrelated absorption \citep{Pierietal2014}. To get this pseudo-continuum, we perform a simple spline fit of the stack.

The final composite spectrum is shown in Figure~\ref{fig:CompositeCombinedCatalogue}.
The purity estimate we obtain from the Lyman-8 flux decrement is 76.3\%. 
We repeat this analysis on the DLA sample from the DLA Toolkit only (Section~\ref{subsect:dr2_catalog}), which can be seen in Appendix \ref{appx:DLAToolkitOnlyComposite}.

\section{Future Outlook}
\label{sect:future}
The impact of DLA masking on BAO using the various DLA catalogs presented in this work is explored by \citet{casas_mocks}. 
The authors demonstrate that BAO parameters remain highly robust against minor variations in the $N_{\texttt{HI}}$ threshold, purity, and completeness. For example, they show that the individual catalogs from the DLA Toolkit, GP, and CNN methods as well as the final combined catalog produce consistent BAO results with similar uncertainties (see their Figures 12 and 13). 

While BAO is robust to variation in the DLA catalog, other Ly$\alpha$ forest analyses may benefit from adopting alternative catalog strategies than that presented in Section~\ref{BAO_combined_catalog} to improve performance in the $N_\texttt{HI}$ and S/N extremes. In particular, the Ly$\alpha$ forest full shape, 1-dimensional power spectrum (P1D), and 3-dimensional power spectrum (P3D) analyses are more sensitive to DLA incompleteness than BAO \citep[e.g.][]{mcdonald05,rogers18b}. For example, the convention for Ly$\alpha$ P1D accounts for incompleteness with a systematic error budget on the measurement. In the optimal estimator measurement from DESI DR1 \citep[][]{karacayli25}, DLA incompleteness comprises over 50\% of the total error budget at large scales and at $z \leq 3.2$ with an estimated DLA completeness of 85\%.

In Appendix~\ref{appx:diff_combined_strategy}, we show that relaxing the redshift matching criterion between catalogs can boost completeness at $\log_{10} N_\texttt{HI} > 22.0$ by 30\%. Meanwhile, the performance outside of this regime is relatively stable against this change. While DLAs of this high column density are rare, they do compromise a significant fraction of the forest when present. The completeness at lower $N_\texttt{HI}$ could also be improved by applying the minimum column density cut at a later point in the combined catalog's construction. Since predicted $N_\texttt{HI}$ from the CNN is negatively biased, we considered cutting the combined catalog on predicted $\log_{10} (N_\texttt{HI}) >20.3$ from the GP only. 
This is similar to the approach by \cite{dr1_dlas} and provides a catalog that is 
83.8\% complete (4.8\% higher); however, the purity decreases to 77.8\% (7\% lower). 
The change in these metrics relative to Section~\ref{BAO_combined_catalog} is driven by detections with $20.3 < \log_{10} (N_\texttt{HI}) < 20.5$.

Our purity and completeness analysis for the DLA Toolkit, GP, and CNN illustrates some of the strengths and weaknesses of each algorithm.  As it is a major focus of this paper, we discuss specific modifications to the DLA Toolkit that could improve performance. 

The most notable DLA Toolkit failure mode is the reduced completeness at high S/N (Figure~\ref{fig:snr_purity_completeness}). This is primarily driven by poor $\chi^2$ relaxation on the ($N_\texttt{HI}$, $z$)-surface owing to local minima within true DLA troughs. As such, the DLA Toolkit often returns 2 DLA solutions corresponding to a single true DLA, each with low predicted $N_\texttt{HI}$. Future code releases will particularly target the high S/N regime for improvement. Potential solutions are altering the procedure for $\chi^2$ surface relaxation such that local minima traps are disfavored. For example, the step size of $N_\texttt{HI}$ and $z$ can decrease with increasing SNR to retain an efficient runtime while allowing for a more detailed search at high SNR. Alternatively (or additionally), a minimum $\Delta z$ restriction can be imposed between DLAs on the same sightline to discourage multiple solutions for the same DLA or it can be used as a trigger to refit the parabola with a wider refined search window. 

Another path to enhance performance is replacing the quasar flux model with one trained on DESI quasar spectra. The current quasar model, introduced in Section~\ref{subsec:dlatoolkit}, was trained on SDSS quasars. The bluest wavelengths of SDSS spectra suffer from poor calibration \citep{margala16}; therefore, we may expect degraded modeling performance in exactly the wavelength region in which we are looking for DLAs. Ongoing work related to upgrading the quasar model shows promising results, specifically regarding improved $N_\texttt{HI}$ accuracy.

As for the GP and CNN DLA finders, there is ongoing work to retrain both methods on the higher-quality spectra that are now available with DESI. The CNN model is being reconstructed and retrained on the new generation of DESI mocks that feature improved realism. As for the GP, DESI DR2 data is being used to upgrade the null model with the specific intent of improving its ability to distinguish between DLA and non-DLA sightlines in the low-S/N regime. 

\section{Summary}
\label{sect:discussion}
In this work, we presented the DLA Toolkit software for automated DLA detection. The technique uses a spectral template fitting approach that identifies DLA positions and estimates their column densities via sliding Voigt profiles while accommodating variance in the quasar's flux. We explore the performance of the DLA Toolkit with respect to S/N, $N_\texttt{HI}$, and detection confidence as defined by the $\chi^2$ parameter in Equation~\eqref{eq:chi2} on a sample of simulated mock spectra. The best performance is achieved for S/N~$>2$, and we recommend applying a $\Delta\chi^2>0.03$ cut to enhance purity. With these cuts, the mock DLA sample from the DLA Toolkit is approximately 80\% pure and complete for non-BAL sightlines, and the predicted redshifts are highly accurate. The DLA Toolkit tends to overestimate column densities with an accuracy that generally improves with S/N. We make available a catalog of candidate DLAs found with this technique on DESI DR1.

We combined the DLA catalog from the DLA Toolkit with catalogs from the GP DLA finder by \citet{ho21} and the CNN DLA finder by \citet{wang22} on the DESI DR2 quasar sample from the Ly$\alpha$ forest BAO analysis presented in \cite{DESI_lyaBAO}. Our goal is to construct a DLA catalog for the Ly$\alpha$ BAO analysis that optimally balances purity and completeness, thereby minimizing DLA impact on BAO parameter uncertainty. The combined catalog prescription is presented in Section~\ref{BAO_combined_catalog}, summarized by Equations~\eqref{eq:dlatoolkit_detection}$-$\eqref{eq:combined_cat}. Given the performance of all three methods degrades at low S/N, we restrict our combined catalog to S/N~$>2$ to avoid unnecessary loss of Ly$\alpha$ signal. We additionally require predicted $\log_{10} N_\texttt{HI} > 20.3$ owing to the difficulty of accurately identifying high column density absorbers below this. The DLA parameters of the combined catalog are determined by the GP DLA finder, which provides the highest accuracy predicted $N_\texttt{HI}$ (Figure~\ref{fig:z_nhi_hists}) of the three algorithms. 

An analysis of the combined catalog strategy on mocks suggests the combined DLA catalog for DR2 Ly$\alpha$ BAO is 84.8\% pure and 79.0\% complete when BAL sightlines are excluded. 
Compared to the DLA catalog strategy used for DR1 Ly$\alpha$ BAO, this constitutes an improvement of roughly 18\% in completeness while losing only 5\% in purity. When BAL sightlines are included, the purity estimate drops to 66.4\% with a relatively unaffected completeness of 77.6\%.
Lastly, we obtain an estimate for purity of 76.8\% on real data through spectral stacking of the combined catalog for DR2 Ly$\alpha$ BAO.

\section*{Data Availability}
The DLA catalog from the DLA Toolkit for data collected during DESI DR1 observations is available at
\url{https://data.desi.lbl.gov/public/dr1/vac/dr1/dla-toolkit}. All data shown in figures can be downloaded from \url{https://doi.org/10.5281/zenodo.14948183}.

\begin{acknowledgments}
We thank the anonymous referee for valuable comments that greatly improved the quality of this manuscript.
AB is supported by the U.S. Department
of Energy, Office of Science,
Office of High-Energy Physics under Contract No. DE–AC02–05CH11231.
DMS and MMP acknowledge the support of the French National Research Agency (ANR) under contracts ANR-22-CE31-0009 and ANR-22-CE31-0026.
This material is based upon work supported by the U.S. Department of Energy (DOE), Office of Science, Office of High-Energy Physics, under Contract No. DE–AC02–05CH11231, and by the National Energy Research Scientific Computing Center, a DOE Office of Science User Facility under the same contract. Additional support for DESI was provided by the U.S. National Science Foundation (NSF), Division of Astronomical Sciences under Contract No. AST-0950945 to the NSF’s National Optical-Infrared Astronomy Research Laboratory; the Science and Technology Facilities Council of the United Kingdom; the Gordon and Betty Moore Foundation; the Heising-Simons Foundation; the French Alternative Energies and Atomic Energy Commission (CEA); the National Council of Humanities, Science and Technology of Mexico (CONAHCYT); the Ministry of Science, Innovation and Universities of Spain (MICIU/AEI/10.13039/501100011033), and by the DESI Member Institutions: \url{https://www.desi.lbl.gov/collaborating-institutions}. Any opinions, findings, and conclusions or recommendations expressed in this material are those of the author(s) and do not necessarily reflect the views of the U. S. National Science Foundation, the U. S. Department of Energy, or any of the listed funding agencies.

The authors are honored to be permitted to conduct scientific research on Iolkam Du’ag (Kitt Peak), a mountain with particular significance to the Tohono O’odham Nation.
\end{acknowledgments}

\bibliographystyle{mod-apsrev4-2} 
\bibliography{main}

\appendix
\section{Relaxed Redshift Matching Between Catalogs} 
\label{appx:diff_combined_strategy}
As discussed in Section~\ref{BAO_combined_catalog}, the strategy for combining the DLA catalogs from the DLA Toolkit, GP, and CNN methods results in relatively low completeness for DLAs with $N_{\texttt{HI}} > 22.0$ (see Figure~\ref{fig:combined_matrix}).
One contributing factor is the decision to associate DLA detections from different methods only if they fall within 800 km~s$^{-1}$ of each other, following \cite{DESI-Y1KP6, dr1_dlas}.

\begin{figure*}[t]
\centering
    \includegraphics[width=\columnwidth]{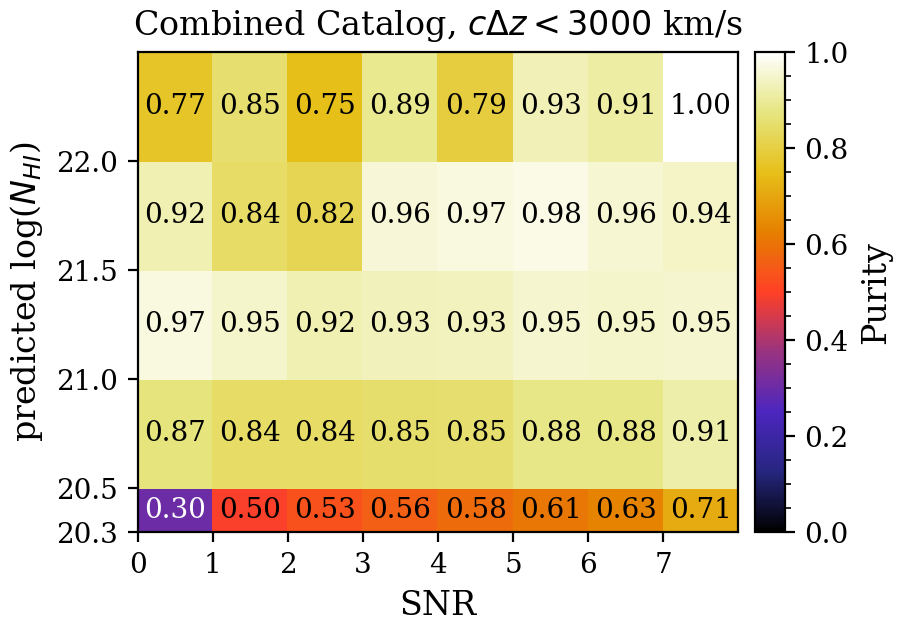}
    \includegraphics[width=\columnwidth]{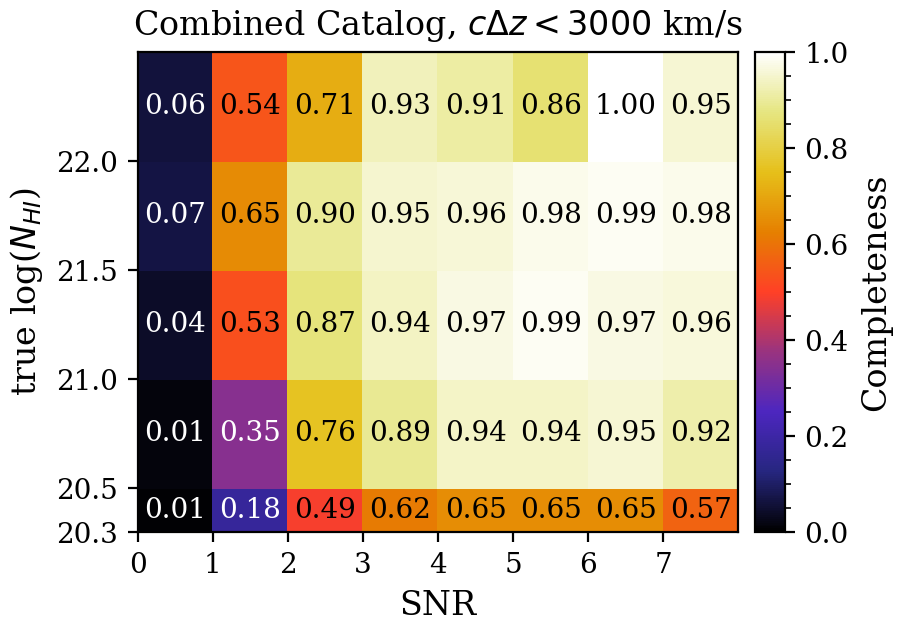}
    \caption{
        The purity (top) and completeness (bottom) of the combined catalog when relaxing redshift matching criteria to within 3,000 km~s$^{-1}$ as function of S/N and $\log_{10}(N_{\texttt{HI}})$.
        The $6<$~S/N~$<7$, $\log_{10} (N_\texttt{HI}) > 22.0$ bin with 100\% completeness contains only 6 DLAs.}
    \label{fig:appx-matrix-low_tol}
\end{figure*} 

To explore an alternative approach, we created a combined catalog in which DLA detections from different methods are associated if they fall within 3,000 km~s$^{-1}$, similar to the threshold used to define true detections in Equation~\eqref{eq:detection_def}. 
We then applied the same combined catalog definition from Equation~\eqref{eq:combined_cat} to assess the impact of this higher tolerance on purity and completeness.

The resulting purity and completeness, evaluated in the same S/N and $N_{\texttt{HI}}$ bins as in Section~\ref{sect:performance}, are shown in Figure~\ref{fig:appx-matrix-low_tol}. 
With the increased association tolerance, completeness in all the $\log_{10}(N_{\texttt{HI}}) > 22.0$ bins improves by approximately 30\% when S/N$ > 2$, while purity decreases by less than 10\% and actually increases for the highest S/N. 
Since high $N_{\texttt{HI}}$ systems are rare, the overall impact on catalog purity and completeness is minimal. 
The total purity and completeness are now 83.9\% and 80.0\%, respectively -- changing by roughly 1\% compared to the original catalog -- while significantly improving completeness at high $N_{\texttt{HI}}$, which is particularly relevant because these systems compromise a significant fraction of the forest when present.

\section{DLA Toolkit DR2 Catalog Composite Spectrum} 
\label{appx:DLAToolkitOnlyComposite}
\begin{figure*}[t]
    \centering
    \includegraphics[width=1.0\linewidth]{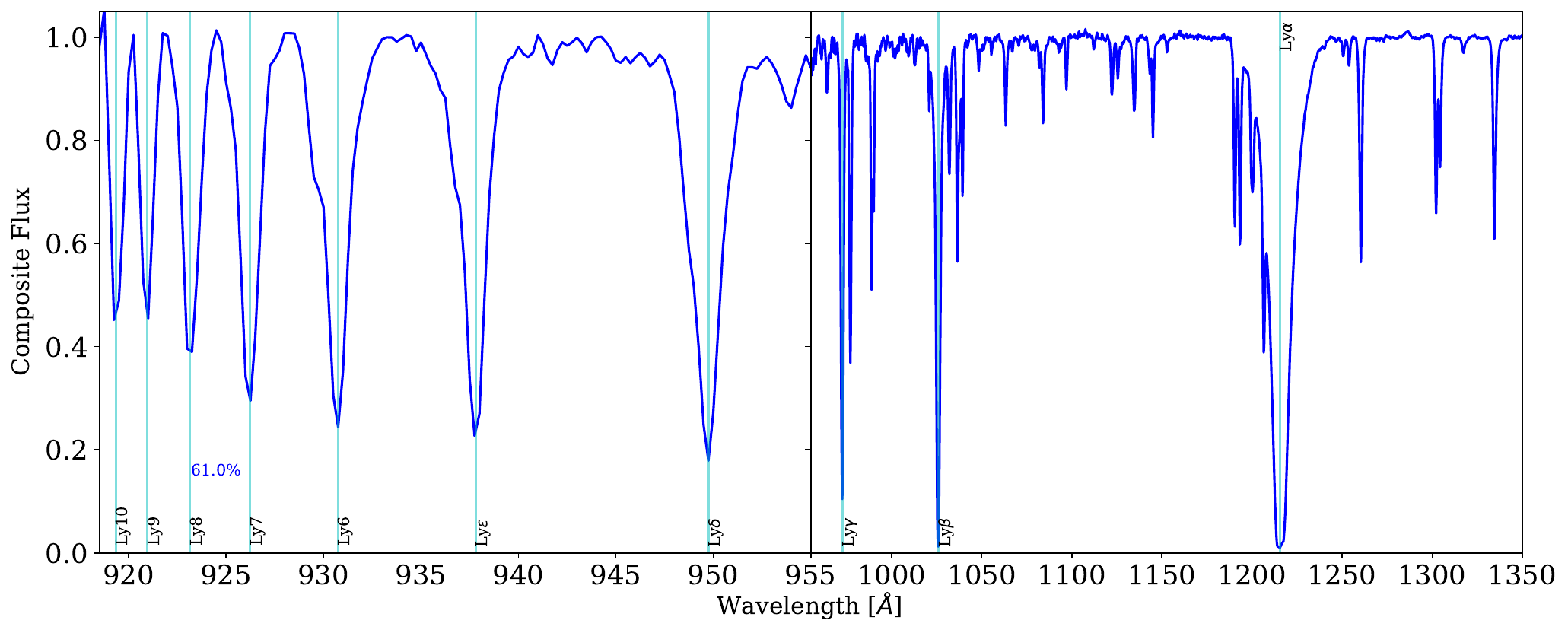}
    \caption{Composite spectrum of DLAs in the DR2 catalog from DLA Toolkit, according to the properties defined in Appendix \ref{appx:DLAToolkitOnlyComposite}. Lyman series lines are marked with vertical cyan lines. The percentage close to the Lyman-8 line represents the purity of the selected DLA sample for DLAs with predicted $z\gtrsim2.9$.}
    \label{fig:TemplateComposite}
\end{figure*}

Calculating purity from a composite spectrum is a tool we can use not only for the combined catalog (see Section~\ref{subsubsect:purity_validation}), but for all the DLA finding methods. Here we tackle the DLAs found by the DLA Toolkit exclusively to understand what data tells us about purity with this method. We select candidate DLAs with $\Delta \chi_r^2 > 0.03$, S/N~$> 2$, and predicted $\log_{10} (N_\texttt{HI}) > 20.3$ from the DESI DR2 DLA Toolkit catalog, while removing BAL sightlines. We follow the same procedure outlined in Section~\ref{subsubsect:purity_validation} to construct a composite spectrum from the selected sample and obtain a purity estimate. 

As noted in Section~\ref{subsubsect:purity_validation}, since only true DLAs will exhibit 0\% transmission at higher order Lyman series wavelengths, the flux decrement at these wavelengths acts as a proxy for sample purity. In this work, we choose the Lyman-8 line for this measurement. Thus, candidate DLAs must have a minimum predicted redshift $z\gtrsim2.9$ to contribute to the stack owing to the wavelength coverage of the DESI spectrographs. This corresponds to $\sim$15\% (5,498 candidate DLAs) of the DR2 DLA Toolkit catalog after applying the aforementioned quality cuts.

The resulting DLA composite spectrum is shown in Figure~\ref{fig:TemplateComposite}. The Lyman-8 flux deficit provides a purity estimate of roughly $61\%$.  As a reminder, the purity measured here can be considered a conservative estimate for the entire sample, owing to the higher forest opacity increasing the frequency of DLA interlopers with higher DLA redshifts. For comparison, we measure a purity of $82$\% from the mock DLA sample with predicted $z>2.9$.

An intriguing feature seen in the stack is the non-zero Ly$\alpha$ trough (offset by $\sim0.025)$, hinting at where the DLA Toolkit may under-perform. Perhaps there are interlopers with similar absorption profiles as DLAs for which the Voigt profile addition in the model reduces the $\chi_r^2$ sufficiently to constitute a detection. Fortunately, the final combined catalog strategy appears appears robust against this type of contamination, as even with the lower purity seen here, the combination of GP, CNN, and DLA Toolkit produces significantly high purity.

\end{document}